\documentclass[preprint,tightenlines,aps,prd,showpacs,nofootinbib]{revtex4}
\usepackage{graphicx}  
\usepackage{amsmath}
\usepackage{amssymb}
\usepackage{bm}
\begin{document}
%%%%%%%%%%%%%%%%%%%%%%%%%%%%%%%%%%%%%%%%%%%%%%%%%%%%%%%%%%%%%%%%%%%%%%
\title{\mbox{}\\[10pt] Analysis of $\bm{J/\psi \, \pi^+ \pi^-}$
                     and $\bm{D^{0} {\bar D}^0 \pi^0}$ Decays 
                      of the $\bm{X(3872)}$}
%%%%%%%%%%%%%%%%%%%%%%%%%%%%%%%%%%%%%%%%%%%%%%%%%%%%%%%%%%%%%%%%%%%%%%
\author{Eric Braaten}
%\email{braaten@pacific.mps.ohio-state.edu}
\author{James Stapleton}
%\email{meng@pacific.mps.ohio-state.edu}
\affiliation{Physics Department, Ohio State University, Columbus, Ohio
  43210, USA}
\date{\today}
%%%%%%%%%%%%%%%%%%%%%%%%%%%%%%%%%%%%%%%%%%%%%%%%%%%%%%%%%%%%%%%%%%%%%%
\begin{abstract}
We analyze recent data from the Babar and Belle Collaborations 
on the $X(3872)$ resonance in  the $J/\psi \, \pi^+ \pi^-$ 
and $D^0 \bar{D}^0 \pi^0$ decay channels, taking careful account
of the universal features of an S-wave threshold resonance.
Because the line shapes for such a resonance are not integrable 
functions of the energy, the resonance parameters depend on 
the prescriptions used to define them.
In recent experimental analyses of the $D^0 \bar{D}^0 \pi^0$ channel, 
an event near the $D^{*0} \bar{D}^0$ threshold was assumed 
to come from $D^{*0} \bar{D}^0$ or $D^{0} \bar{D}^{*0}$ 
and was therefore assigned an energy above the threshold.  
Taking this effect into account, our analysis of the
$D^0 \bar{D}^0 \pi^0$ data gives a mass 
for the $X(3872)$ that is below the $D^{*0} \bar{D}^0$ threshold.
Our analyses in both the $J/\psi \, \pi^+ \pi^-$ and  
$D^0 \bar{D}^0 \pi^0$ channels are consistent with the identification 
of the  $X(3872)$ as an extremely weakly-bound charm meson molecule.
\end{abstract}
% 12.38.-t   Quantum chromodynamics
% 12.39.St  Factorization
% 13.20.Gd  Decays of J/psi, Upsilon, and other quarkonia
% 14.40.Gx   Mesons with S=C=B=0, mass > 2.5 GeV (including quarkonia)

%\keywords{}

%%%%%%%%%%%%%%%%%%%%%%%%%%%%%%%%%%%%%%%%%%%%%%%%%%%%%%%%%%%%%%%%%%%%%%
\maketitle


\section{Introduction}

In August 2003, the Belle Collaboration discovered a
new $c \bar c$ meson that they named the $X(3872)$ \cite{Choi:2003ue}.
This marked the beginning of a new era in $c \bar c$ meson
spectroscopy in which discoveries at the $B$ factories 
have tripled the number of known $c \bar c$ mesons above
the charm meson pair threshold.  Of these new $c \bar c$ mesons,
the $X(3872)$ remains the one for which by far the most 
experimental information is available.  Still the nature 
of this state is not universally recognized in the 
high energy physics community.

There are two crucial pieces of experimental information 
that determine the nature of the $X(3872)$ unambiguously.
One is its mass as measured in the $J/\psi \, \pi^+ \pi^-$ decay mode.
By combining the most recent measurements by the Belle, Babar, 
and CDF Collaborations \cite{Belle:2008te,Aubert:2008gu,Aaltonen:2009vj}, 
its mass is determined to be $M_X = 3871.55 \pm 0.20$ MeV.
This mass is extremely close to the $D^{*0}\bar{D}^0$ threshold.
The energy relative to the threshold is
$-0.25 \pm 0.40$ MeV.  The central value corresponds 
to a bound state with binding energy $E_X = 0.25$ MeV.
The second crucial piece of information is the $J^{PC}$ quantum numbers.
Observations of decays into $J/\psi \, \gamma$ and
$\psi(2S) \, \gamma$ by the Belle and 
Babar Collaborations \cite{Abe:2005ix,Aubert:2006aj,Babar:2008rn}
imply that $X$ is even under charge conjugation.
The spin and parity quantum numbers have been constrained 
by the Belle and CDF Collaborations \cite{Abe:2005iya,Abulencia:2006ma}
from studies of the angular distributions in $J/\psi \, \pi^+ \pi^-$ decays.
The CDF analysis is compatible only with 
$J^{PC} = 1^{++}$ and $2^{-+}$ \cite{Abulencia:2006ma}. 
The possibility $2^{-+}$ is disfavored by the observation 
of the decay into $\psi(2S) \, \gamma$ \cite{Babar:2008rn}, 
because it would have to overcome multipole suppression.
The possibility $2^{-+}$ is also disfavored by the observation 
of decays into $D^{*0}\bar{D}^0$ by the Belle and Babar
Collaborations \cite{Gokhroo:2006bt,Babar:2007rva,Belle:2008su}, 
because it would have to overcome angular-momentum suppression
associated with the tiny energy relative to the 
$D^{*0}\bar{D}^0$ threshold.  We will assume from now on that
the quantum numbers of the $X(3872)$ are $1^{++}$.

Given that its quantum numbers are $1^{++}$,
the $X(3872)$ has an S-wave coupling to the
charm meson pairs $D^{*0}\bar{D}^0$ and $D^0\bar{D}^{*0}$.
The closeness of the mass to the $D^{*0} \bar D^0$ threshold
implies that it is a resonant coupling.
This state is therefore governed by the universal properties 
of S-wave threshold resonances that are predicted by 
nonrelativistic quantum mechanics \cite{Braaten:2004rn}.
We can conclude that the $X(3872)$ is a charm meson molecule whose
constituents are a superposition of $D^{*0}\bar{D}^0$ 
and $D^0\bar{D}^{*0}$.  
Among the universal properties of this molecule is that the 
root-mean-square separation of its constituents is $\sqrt{\mu E_X/2}$, 
where $\mu$ is the reduced mass of the $D^{*0}\bar{D}^0$.  
The tiny binding energy $E_X$ implies a large RMS separation,
with the central value 0.25~MeV corresponding to an 
astonishing RMS separation of about 6 fermis.

Some of the confusion regarding the nature of the $X(3872)$ 
has been prompted by measurements of the 
$D^0 \bar{D}^0 \pi^0$ and $D^0 \bar{D}^0 \gamma$ decay modes
\cite{Gokhroo:2006bt,Babar:2007rva,Belle:2008su}.
In the most recent analyses by the Babar and Belle Collaborations, 
these decay modes have been analyzed as if they were decays into
$D^{*0}\bar{D}^0$ and $D^0\bar{D}^{*0}$
\cite{Babar:2007rva,Belle:2008su}.
The resulting energy distribution must by definition vanish below the 
$D^{*0}\bar{D}^0$ threshold and it has a peak just above the threshold.
Measurements of the position and width of this peak have been interpreted 
incorrectly as measurements of the mass and width of the $X(3872)$.
For example, in the 2008 edition of the Review of Particle Physics 
\cite{Amsler:2008zzb}, the Particle Data Group determined their 
average for the mass of the $X(3872)$ by combining four values 
below the threshold from $J/\psi \, \pi^+ \pi^-$ decays
with two values above the threshold from $D^0 \bar{D}^0 \pi^0$  
and $D^0 \bar{D}^0 \gamma$ decays. In the PDG average mass, 
the 3.5 sigma discrepancy between these two sets of
measurements was taken into account 
by increasing the error by a scale factor of 2.5.
The Particle Data Group also took as their average for the decay width 
of the $X(3872)$ the width of the $D^{*0}\bar{D}^0$ energy distribution 
measured by the Babar Collaboration.

Several authors have misinterpreted the measurements 
of the $D^0 \bar{D}^0 \pi^0$ decay modes as evidence that the $X(3872)$ 
is not a bound state with mass below the $D^{*0}\bar{D}^0$
threshold but instead as a ``virtual state'' which is unbound
\cite{Hanhart:2007yq,Voloshin:2007hh,Zhang:2009bv}.
The signature for a virtual state associated with an S-wave 
threshold resonance is an enhancement in the production of 
$D^{*0}\bar{D}^0$ just above the threshold together with the 
absence of a resonance  in $D^0 \bar{D}^0 \pi^0$ below the threshold.
This should be contrasted with a bound state, whose signature is
a similar enhancement above the threshold together with
a resonance below the threshold.
In misinterpreting the $D^0 \bar{D}^0 \pi^0$ data 
as evidence for a virtual state, the authors of
Refs.~\cite{Hanhart:2007yq,Voloshin:2007hh,Zhang:2009bv}
did not take into account that a bound state of $D^{*0}\bar{D}^0$ 
can decay into $D^0 \bar{D}^0 \pi^0$ and $D^0 \bar{D}^0 \gamma$
through decays of its constituent $D^{*0}$ or $\bar D^{*0}$. 
This conceptual error was pointed out in Ref.~\cite{Braaten:2007dw},
and an analysis that takes proper account of the bound state 
was carried out. 

In this paper, we carry out analyses of the recent 
data from the Babar and Belle Collaborations on the $X(3872)$ resonance
in the $J/\psi \, \pi^+ \pi^-$ and $D^0 \bar{D}^0 \pi^0$ decay channels.  
We begin in Section~\ref{sec:LS-short} by describing the line shape 
of an S-wave threshold resonance in a short-distance decay channel, 
such as the $J/\psi \, \pi^+ \pi^-$ decay mode of $X(3872)$.
In Section~\ref{sec:psipipi}, we use that line shape to analyze
the most recent data from the Belle and Babar Collaborations 
on the $J/\psi \, \pi^+ \pi^-$ decay channel.
We proceed in Section~\ref{sec:LS-long} to describe the line shape 
of $X(3872)$ in the $D^0 \bar{D}^0 \pi^0$ decay channel,
which involves the decay of a constituent.
In Section~\ref{sec:DDpi}, we use that line shape to analyze
the most recent data from the Belle and Babar Collaborations 
on the $D^0 \bar{D}^0 \pi^0$ decay channel.
We take into account the experimental procedure that identifies
$D^0 \bar{D}^0 \pi^0$ events near the $D^{*0} \bar{D}^0$ threshold
as $D^{*0} \bar{D}^0$ or $D^{0} \bar{D}^{*0}$ events
above the threshold.  Our analysis of these energy distributions, 
which are nonzero only above the $D^{*0}\bar{D}^0$ threshold, 
gives a mass for the $X(3872)$ that is below the threshold.
In Section~\ref{sec:critique}, 
we present a critical discussion of previous theoretical analyses 
of the line shapes of the $X(3872)$.
Our results are summarized in Section~\ref{sec:summary}.

\section{Line shape in the $\bm{J/\psi \, \pi^+ \pi^-}$ decay channel}
\label{sec:LS-short}

If there is an S-wave resonance very close to the threshold 
for a pair of particles with short-range interactions, their
scattering length $a$ is large compared to the range of their interaction.
The line shapes associated with an S-wave threshold resonance 
have some unusual features that are not ordinarily encountered in 
high energy physics. The line shapes for
decay modes that involve the decay of a constituent are different 
from those for all other decay modes, 
because they can proceed even when the constituents 
have a large separation of order $a$.  For all other decay modes,
the constituents must approach to within a much smaller distance 
comparable to the range of the interaction.
We will refer to these two classes of decay modes as 
{\it constituent decay modes} and {\it short-distance decay modes}, 
respectively.
In the case of the $X(3872)$, the constituent decay modes are
$D^0 \bar{D}^0 \pi^0$ and $D^0 \bar{D}^0 \gamma$
and an example of a short-distance decay mode is $J/\psi \, \pi^+ \pi^-$. 
In this section, we summarize the essential aspects of the line shape 
for short-distance decay modes.

The line shape for a resonance near a scattering threshold
is proportional to $|f(E)|^2$, where $f(E)$ is the analytic continuation
of the scattering amplitude in the total energy $E$ of the particles
in their center-of-mass frame.
The universal scattering amplitude for an 
S-wave threshold resonance has the form
\begin{equation}
f(E) = \frac{1}{- \gamma
                + \sqrt{-2 \mu (E + i \epsilon)}}~,
\label{Eq:Universal}
\end{equation}
where $E$ is the energy relative to the threshold, $\mu$ is the 
reduced mass, and $\gamma=1/a$ is the inverse scattering length.
In the case of the $X(3872)$ resonance, the relevant scattering 
amplitude is for $D^{*0}\bar{D}^0$ mesons in the $1^{++}$ channel. 
For quantitative applications, the scattering amplitude 
in Eq.~(\ref{Eq:Universal}) must be modified to take into account the 
nonzero width of the $D^{*0}$ and the existence of inelastic 
scattering channels for the charm mesons \cite{Braaten:2007dw}.
By analytically continuing the parameters in 
Eq.~(\ref{Eq:Universal}) to complex values,
we obtain
\begin{equation}
f(E) = \frac{1}{- (\gamma_{\rm re} + i \gamma_{\rm im}) 
                + \sqrt{-2 \mu (E + i \Gamma_{*0}/2)}}~,
\label{Eq:ScatteringAmplitude}
\end{equation}
where $\mu = 966.6$ MeV is the reduced mass of the $D^{*0}$ and $\bar{D}^0$,
$\Gamma_{*0} = 65.5 \pm 15.4$ keV is the total width of the $D^{*0}$,
and $\gamma_{\rm re} + i \gamma_{\rm im}$ 
is the complex inverse scattering length.
The effects of the decays of $D^{*0}$ into $D^0 \pi^0$ and $D^0 \gamma$
are taken into account through $\Gamma_{*0}$.
The effects of the inelastic scattering channels 
for $D^{*0} \bar{D}^0$, such as $J/\psi \, \pi^+ \pi^-$,
are taken into account through $\gamma_{\rm im}$, which must be positive.
The scattering amplitude in Eq.~(\ref{Eq:ScatteringAmplitude})
should be accurate as long as the energy is within about an MeV of the
threshold.

For a short-distance decay channel, the only dependence of the 
line shape on the energy $E$ is from the resonance factor $|f(E)|^2$.
If $\gamma_{\rm re}$ is positive, the line shape $|f(E)|^2$
has a resonance peak below the $D^{*0} \bar{D}^0$ threshold.
Defining the binding energy and decay width
for this resonance is problematic, because the line shape is 
not that of a conventional Breit-Wigner resonance.
Our prescriptions for the binding energy $E_X$ 
and the width $\Gamma_X$ are that the pole of the amplitude $f(E)$ 
in the complex energy $E$ is at $-E_X - i \Gamma_X/2$:
\begin{subequations}
\begin{eqnarray}
E_X &\equiv&   \frac{\gamma_{\rm re}^2 - \gamma_{\rm im}^2}{2 \mu}, 
\label{Eq:EX}
\\
\Gamma_X &\equiv& \Gamma_{*0} + \frac{2 \gamma_{\rm re} \gamma_{\rm im}}{\mu}~.
\label{Eq:GamX}
\end{eqnarray}
\label{Eq:EX-GamX}
\end{subequations}
In the case $\Gamma_X \ll 2 E_X$, the shape of the resonance is 
approximately that of a nonrelativistic 
Breit-Wigner resonance in the region $|E + E_X| \ll E_X$.  Its peak is
at $-E_X$ and its full width at half maximum is $\Gamma_X$,
justifying the interpretation of $E_X$ and $\Gamma_X$ as the binding energy 
and decay width of the resonance.
If $\Gamma_X /(2 E_X)$ is not small, the variables $E_X$ and $\Gamma_X$
defined by Eqs.~(\ref{Eq:EX-GamX}) have no 
precise physical interpretations. 

If $\gamma_{\rm re}$ is negative, the line shape $|f(E)|^2$  
has a peak very near the $D^{*0}\bar{D}^0$ threshold.  
In the limit $\Gamma_{*0} \to 0$, the peak is a cusp 
with a discontinuity in the slope that arises from the square root in 
Eq.~(\ref{Eq:ScatteringAmplitude}).
The effect of the $D^{*0}$ width is to smooth out the cusp.
In this case, the variables $E_X$ and $\Gamma_X$
defined by Eqs.~(\ref{Eq:EX-GamX}) specify the location of a pole 
on the second sheet of the complex energy $E$.  Thus they have no simple
physical interpretations.
 
The binding energy $E_X$ and the width $\Gamma_X$ can not be measured directly,
because they are defined in terms of the analytic continuation 
of the scattering amplitude $f(E)$ to complex values of the energy $E$.
An alternative pair of variables that can in principle be measured directly 
are the position $E_{\rm max}$ of the peak in the line shape
and its full width at half-maximum $\Gamma_{\rm fwhm}$.
The position $E_{\rm max}$ of the peak satisfies
\begin{eqnarray}
2  \mu E_{\rm max} &+& \gamma_{\rm re} 
\left( \mu \sqrt{E_{\rm max}^2 + \Gamma_{*0}^2/4} - \mu E_{\rm max} \right)^{1/2}
\nonumber \\
&+& \gamma_{\rm im} 
\left( \mu \sqrt{E_{\rm max}^2 + \Gamma_{*0}^2/4} + \mu E_{\rm max} \right)^{1/2}
= 0~.
\label{Eq:eqEmax}
\end{eqnarray}
The full width of the line shape at half-maximum is given by
$\Gamma_{\rm fwhm} = E_+ - E_-$, where $E_\pm$ are the two solutions of
\begin{equation}
|f(E_\pm)|^2 = \frac12 |f(E_{\rm max})|^2.
\label{Eq:eqGammafwhm}
\end{equation}
If $\gamma_{\rm re}>0$, 
the solutions for $E_{\rm max}$, $E_+$, and $E_-$
can be expanded in powers of $\Gamma_{*0}$.
The expansions for $E_{\rm max}$ and $\Gamma_{\rm fwhm}$ are
\begin{subequations}
\begin{eqnarray}
E_{\rm max} &=& 
- \frac{1}{2 \mu} 
\left( \gamma_{\rm re}^2
+ \frac{\gamma_{\rm im}}{\gamma_{\rm re}} (\mu \Gamma_{*0})
+ \frac{\gamma_{\rm re}^2 - 3 \gamma_{\rm im}^2}{4 \gamma_{\rm re}^4} 
	(\mu \Gamma_{*0})^2
 + \ldots \right)~,
\label{Eq:Emax-gamma}
\\
\Gamma_{\rm fwhm} &=& 
\frac{1}{2\mu}
\left( 4 \gamma_{\rm re} \gamma_{\rm im}
+ 2 (\mu \Gamma_{*0})
+ \frac{\gamma_{\rm im}^3 (3 \gamma_{\rm re}^2 - \gamma_{\rm im}^2)}
      {\gamma_{\rm re}^3 (\gamma_{\rm re}^2 - \gamma_{\rm im}^2)^2}
      (\mu \Gamma_{*0})^2 + \ldots \right).
\label{Eq:Gammafwhm-gamma}
\end{eqnarray}
\label{Eq:Emax-Gammafwhm}
\end{subequations}

The normalization of the line shape of $X$ in a short-distance decay mode $F$
produced by the decay of $B^+$ into $K^+ + X$ is proportional to the product 
of the branching fractions for $B^+ \to K^+ + X$ and $X \to F$.
It is convenient to introduce a compact notation for the product of 
these two branching fractions:
\begin{equation}
({\cal B} {\cal B})_F \equiv
{\cal B}[B^+ \to K^+ + X] \, {\cal B}[X \to F] ~.
\label{Eq:BrBr}
\end{equation}
Defining these branching fractions is problematic, because the line shape 
for an S-wave threshold resonance is not an integrable function.
Since $|f(E)|^2$ decreases as $1/|E|$ for large $|E|$,
the integral of $|f(E)|^2$ over $E$ depends logarithmically 
on the endpoints.  
This implies that this product of branching fractions 
cannot be defined uniquely in terms of an integral over the line shape.
The numerical value of $({\cal B} {\cal B})_F$ will inevitably depend 
on the prescription used to define it.
Our prescription is that the normalized line shape 
for $B^\pm \to K^\pm + F$ is
\begin{equation}
\frac{d\Gamma}{dE} \equiv
\Gamma[B^+] \, ({\cal B} {\cal B})_F \,\frac{d \hat \Gamma_{SD}}{dE}~,
\label{Eq:dGdE}
\end{equation}
where the energy-dependent factor is 
\begin{equation}
\frac{d \hat \Gamma_{SD}}{dE} =
\frac{\mu^2 \Gamma_X}{2 \pi (\gamma_{\rm re}^2 + \gamma_{\rm im}^2)} \,
|f(E)|^2~.
\label{Eq:dGhatdE}
\end{equation}
In the case $\Gamma_X \ll 2 E_X$, this line shape 
in the region $|E + E_X| \ll E_X$ is well approximated by a
Breit-Wigner resonance.  The integral of $d \hat \Gamma_{SD}/dE$
over this region is approximately 1,
justifying the interpretation of $({\cal B} {\cal B})_F$ as the 
product of the branching fractions for $B^+ \to K^+ + X$ and $X \to F$.
If $\Gamma_X /(2 E_X)$ is not small, the constant $({\cal B} {\cal B})_F$
defined by Eq.~(\ref{Eq:dGdE}) has no 
precise physical interpretation.  It is simply a convenient variable
for specifying the normalization of the line shape.
An alternative prescription for $({\cal B} {\cal B})_F$ could be obtained
by integrating both sides of Eq.~(\ref{Eq:dGdE}) over a chosen interval 
of the energy $E$ in the threshold region, such as $-2 E_X$ to 0.  
However the numerical value of $({\cal B} {\cal B})_F$ would depend 
on the choice of the endpoints of the interval.

Although the product of branching fractions
depends on the prescription, the ratio of $({\cal B} {\cal B})_F$
for two short-distance decay modes $F$ is independent of the  prescription.
Choosing one of the final states to be $J/\psi \, \pi^+ \pi^-$
and using Eq.~(\ref{Eq:BrBr}), the ratio is
\begin{equation}
\frac{({\cal B} {\cal B})_{F}}{({\cal B} {\cal B})_{J/\psi \, \pi^+ \pi^-}} =
\frac{{\cal B}[X \to F]}{{\cal B}[X \to J/\psi \, \pi^+ \pi^-]}~.
\label{Eq:Bratio}
\end{equation}
The ratio on the right side of Eq.~(\ref{Eq:Bratio}) is the conventional 
branching ratio for decays of $X$ into those states.
This ratio is well-defined for any 
short-distance decay mode $F$, despite the fact that 
a prescription is required to define the products of branching fractions
on the left side of Eq.~(\ref{Eq:Bratio}).

\section{Analysis of the $\bm{J/\psi \, \pi^+ \pi^-}$ decay channel}
\label{sec:psipipi}

%%%%%%%%%%%%%%%%%%%%%%%%%%%%%%%%%%%%%%%%%%%%%%%%%%%%%%%%%%%%%%%%%%%%%%%%%%%%%%%%%%%%%%%%%
\begin{figure}[t]
\includegraphics[width=0.8\textwidth,clip=true]{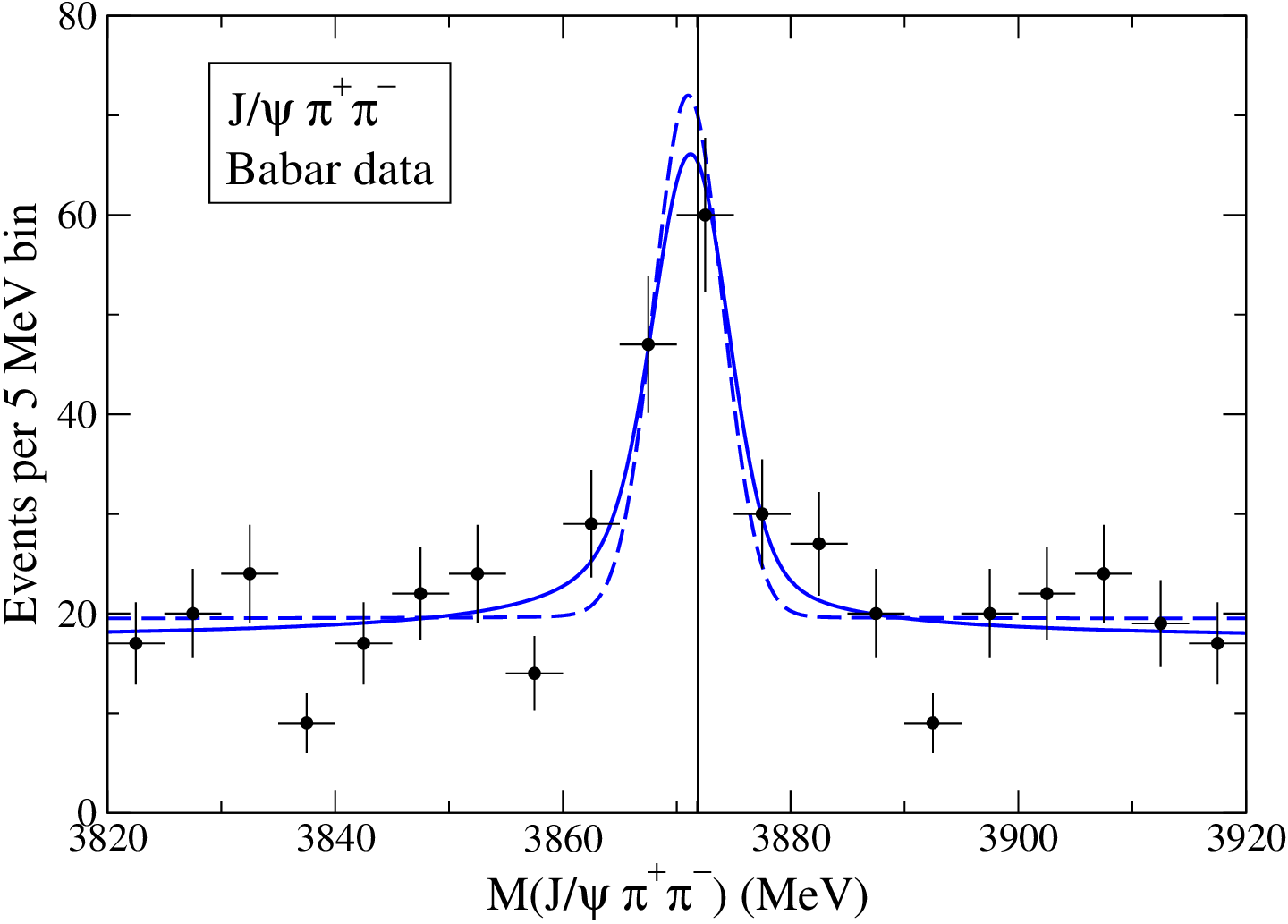}
\caption{
Invariant mass distribution for the $J/\psi \, \pi^+ \pi^-$ decay channel
measured by the Babar Collaboration \cite{Aubert:2008gu}.
The data are the number of events per 5~MeV bin.
The inverse scattering lengths $\gamma_{\rm re} + i \gamma_{\rm im}$
for the two fits are 38.8 MeV (dashed line)
and $(13.6 + 15.5 i)$~MeV (solid line).
The vertical line is the assumed $D^{*0} \bar{D}^0$ threshold
at 3871.8~MeV.
\label{Fig:psipipiBabar}}
\end{figure}
%%%%%%%%%%%%%%%%%%%%%%%%%%%%%%%%%%%%%%%%%%%%%%%%%%%%%%%%%%%%%%%%%%%%%%%%%%%%%%%%%%%%%%%%%

%%%%%%%%%%%%%%%%%%%%%%%%%%%%%%%%%%%%%%%%%%%%%%%%%%%%%%%%%%%%%%%%%%%%%%%%%%%%%%%%%%%%%%%%%
\begin{figure}[t]
\includegraphics[width=0.8\textwidth,clip=true]{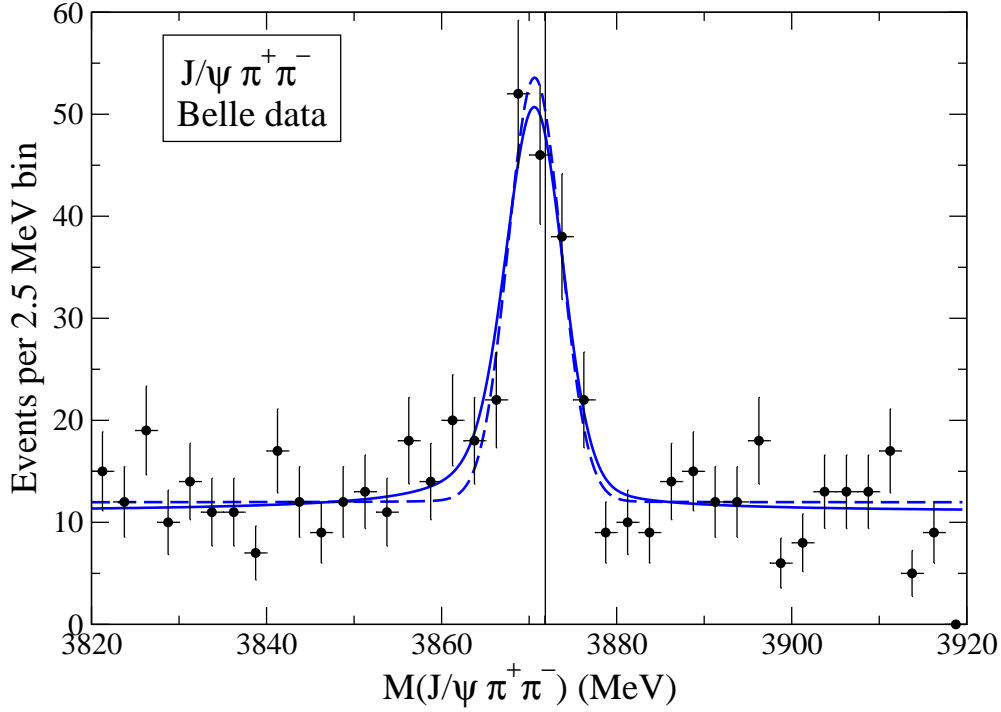}
\caption{
Invariant mass distribution for the $J/\psi \, \pi^+ \pi^-$ decay channel
measured by the Belle Collaboration \cite{Belle:2008te}.
The data are the number of events per 2.5~MeV bin.
The inverse scattering lengths $\gamma_{\rm re} + i \gamma_{\rm im}$
for the two fits are 47.5 MeV (dashed line)
and $(38.4 + 12.0 i)$~MeV (solid line).
The vertical line is the assumed $D^{*0} \bar{D}^0$ threshold
at 3871.8~MeV.
\label{Fig:psipipiBelle}}
\end{figure}
%%%%%%%%%%%%%%%%%%%%%%%%%%%%%%%%%%%%%%%%%%%%%%%%%%%%%%%%%%%%%%%%%%%%%%%%%%%%%%%%%%%%%%%%%

In this section, we analyze recent data from the Belle and Babar Collaborations
on the line shape of the $X(3872)$ in the $J/\psi \, \pi^+ \pi^-$ decay mode
\cite{Belle:2008te,Aubert:2008gu}.
We consider the invariant mass distribution for $J/\psi \, \pi^+ \pi^-$
in the interval from 3820 MeV to 3920 MeV.
For our two data samples, the total number $N_{B \bar B}$ of $B^+ B^-$ 
and $B^0 \bar B^0$ events accumulated
and the number of candidate events for the decay of $B^\pm$ into 
$K^\pm + J/\psi \, \pi^+ \pi^-$ are as follows:
\begin{itemize}
\item
Babar Collaboration \cite{Aubert:2008gu}:
$N_{B \bar B} = 4.55 \times 10^8$, 
471 events in 20 bins of width 5 MeV,
\item
Belle Collaboration \cite{Belle:2008te}:
$N_{B \bar B} = 6.57 \times 10^8$, 
606 events in 40 bins of width 2.5 MeV.
\end{itemize}
The data are shown in Figs.~\ref{Fig:psipipiBabar} and \ref{Fig:psipipiBelle}.
The vertical error bar in a bin with $n$ events is $\sqrt{n}$.
The horizontal error bar indicates the width of the bin.

We take the theoretical line shape for the energy $E$ of 
$J/\psi \, \pi^+ \pi^-$ relative to the $D^{*0} \bar{D}^0$ threshold
to be given by Eqs.~(\ref{Eq:dGdE}) and (\ref{Eq:dGhatdE}).
This line shape will be accurate within about an MeV of the threshold.
We assume that the dominant contributions to the signal come from this 
threshold region.  If this is the case, then a
line shape that remains accurate over a larger energy interval
would give a better approximation only to contributions that are negligible.
To obtain a line shape that remains accurate within about 10~MeV of the
threshold, it is necessary to take into account the effects of
the charged charm meson pairs $D^{*+} D^-$ and $D^{+} D^{*-}$, 
as discussed in Section~\ref{sec:critique}.

To predict the number of events in a given bin of invariant mass,
we need to take into account the background and the energy resolution 
of the experiment.  The resolution must be taken into account 
because the line shape varies 
dramatically over an energy scale smaller than the energy resolution.  
The predicted number of $J/\psi \, \pi^+ \pi^-$ events
in an energy bin of width $\Delta$ centered at $E_i$ 
can be expressed as
\begin{equation}
N_i = 2 N_{B \bar B} 
\left[ ({\cal B} {\cal B})_{J/\psi \, \pi^+ \pi^-} \,
\int_{E_i - \Delta/2}^{E_i + \Delta/2} dE' \int_{-\infty}^\infty dE \,
R(E',E) \, \frac{d \hat \Gamma_{SD}}{dE} + C_{\rm bg} \, \Delta \right]~,
\label{Eq:N-psipipi}
\end{equation}
where $C_{\rm bg}$ is the background 
under the line shape $d \hat \Gamma_{SD}/dE$. 
Our invariant mass interval 3820--3920 MeV is narrow enough 
that we take the background term $C_{\rm bg}$ to be a constant 
independent of $E$.
The experimental resolution is taken into account through
the convolution with the Gaussian resolution function: 
\begin{equation}
R(E',E) =
\frac{1}{\sqrt{2 \pi} \sigma} \,
\exp(-(E'-E)^2/(2\sigma^2)) ~.
\label{Eq:RE}
\end{equation}
We follow Ref.~\cite{Zhang:2009bv} in taking the width 
of the Gaussian to be the same energy-independent constant 
for both experiments: $\sigma = 3$ MeV.

%------------------------------------------------------------------------------------
\begin{table}[t]
\begin{tabular}{l|ccc|cccc}
data set & $\gamma_{\rm re}$ & $\gamma_{\rm im}$ & 
           $({\cal B} {\cal B})_{J/\psi \, \pi^+ \pi^-}$ & 
           $-E_X$ & $\Gamma_X$ & $E_{\rm max}$ & $\Gamma_{\rm fwhm}$ \\
\hline
Babar &
$38.8^{+15.0}_{-23.0}$ &         0          & $8.7^{+1.3}_{-1.3}$ & 
$-0.78^{+0.74}_{-0.80}$ & $0.066 \pm 0.015$ & 
$-0.78^{+0.74}_{-0.80}$ & $0.066 \pm 0.015$ \\
Babar &
$13.6^{+18.3}_{-16.9}$ & $15.5^{+\ 5.8}_{-11.2}$ & $12.3^{+1.8}_{-1.7}$ & 
$+0.03^{+0.39}_{-0.57}$  & $0.50^{+0.61}_{-0.63}$  & 
$-0.13^{+0.38}_{-0.55}$ & $0.56^{+0.58}_{-0.40}$ \\
\hline
Belle &
$47.5^{+7.9}_{-9.6}$   &         0          & $9.6^{+1.1}_{-1.0}$ & 
$-1.17^{+0.56}_{-0.55}$ & $0.066 \pm 0.015$ & 
$-1.17^{+0.56}_{-0.55}$ & $0.066 \pm 0.015$ \\ 
Belle &
$38.4^{+\ 9.8}_{-10.9}$   & $12.0^{+4.6}_{-4.8}$ & $11.1^{+1.3}_{-1.2}$ & 
$-0.69^{+0.52}_{-0.57}$ & $1.02^{+0.44}_{-0.47}$ & 
$-0.77^{+0.51}_{-0.57}$ & $1.02^{+0.44}_{-0.47}$ \\
\end{tabular}
\caption{Results of our analyses of the data for 
$B^\pm \to K^\pm + J/\psi \, \pi^+ \pi^-$.
The four rows correspond to analyses using either the  
Babar data \cite{Aubert:2008gu} or the Belle data \cite{Belle:2008te}
and either setting $\gamma_{\rm im} = 0$ 
or using $\gamma_{\rm im}$ as a fitting parameter.  
All entries are in units of MeV, except for 
$({\cal B} {\cal B})_{J/\psi \, \pi^+ \pi^-}$, 
which is in units of $10^{-6}$.}
\label{tab:psipipi}
\end{table}
%------------------------------------------------------------------------------------

We assume that the number of events in each bin of the 
smeared $J/\psi \, \pi^+ \pi^-$ energy $E'$ 
has a Poisson distribution whose mean 
value is given by $N_i$ in Eq.~(\ref{Eq:N-psipipi}).
We fix the $D^{*0} \bar D^0$ threshold at 3871.8~MeV
and the $D^{*0}$ width $\Gamma_{*0}$ at $65.5$~keV.
The fitting parameters are $\gamma_{\rm re}$, $\gamma_{\rm im}$,
$({\cal B} {\cal B})_{J/\psi \, \pi^+ \pi^-}$, and $C_{\rm bg}$.
We determine the best fit to these parameters by maximizing the
likelihood for the observed distribution.  
For both the Belle and Babar data sets, 
we carry out two fits, one with $\gamma_{\rm im} = 0$ and one with
$\gamma_{\rm im}$ as a fitting parameter.  
The results of our four analyses are presented in Table~\ref{tab:psipipi}.
The error bars on $\gamma_{\rm re}$
and $\gamma_{\rm im}$ are determined by varying these parameters while keeping 
$({\cal B} {\cal B})_{J/\psi \, \pi^+ \pi^-}$ 
and $C_{\rm bg}$ fixed at their central values.
For $\gamma_{\rm im} = 0$, the error bars on $\gamma_{\rm re}$
give the interval within which 
log(Likelihood) differs from its maximum value by less than $1/2$.
If $\gamma_{\rm im}$ is treated as a fitting parameter,
the error bars for $\gamma_{\rm re}$ and $\gamma_{\rm im}$ 
specify the smallest rectangle that contains the error ellipse
in which log(Likelihood) differs from its maximum value by less 
than $1/2$. 
The error bars on $({\cal B} {\cal B})_{J/\psi \, \pi^+ \pi^-}$
are determined by varying this parameter and $C_{\rm bg}$ while keeping 
$\gamma_{\rm re}$ and $\gamma_{\rm im}$ fixed at their central values.

In Table~\ref{tab:psipipi}, we also give the calculated values 
of the position of the resonance and its width 
using two different prescriptions for the parameters.
The values of $-E_X$ and $\Gamma_X$ were 
calculated using Eqs.~(\ref{Eq:EX-GamX}).
The values of $E_{\rm max}$ and $\Gamma_{\rm fwhm}$ were obtained 
by solving Eqs.~(\ref{Eq:eqEmax}) and (\ref{Eq:eqGammafwhm}).
The uncertainty of $\pm 0.36$~MeV in the energy of 
the $D^{*0} \bar D^0$  threshold is taken into account 
as an additional statistical error in $-E_X$ and in $E_{\rm max}$.
The uncertainty of $\pm 15.4$~keV in $\Gamma_{*0}$
is taken into account as an additional statistical error 
in $\Gamma_X$ and in $\Gamma_{\rm fwhm}$.
In Table~\ref{tab:psipipi}, there are significant differences 
between the values of $-E_X$ and $E_{\rm max}$ for the fits
in which $\gamma_{\rm im}$ is used as a fitting parameter.
All four fits give values of 
$({\cal B} {\cal B})_{J/\psi \, \pi^+ \pi^-}$
that are consistent to within the errors 
and approximately equal to $10^{-5}$.

%%%%%%%%%%%%%%%%%%%%%%%%%%%%%%%%%%%%%%%%%%%%%%%%%%%%%%%%%%%%%%%%%%%%%%%%%%%%%%%%%%%%%%%%%
\begin{figure}[t]
\includegraphics[width=0.7\textwidth,clip=true]{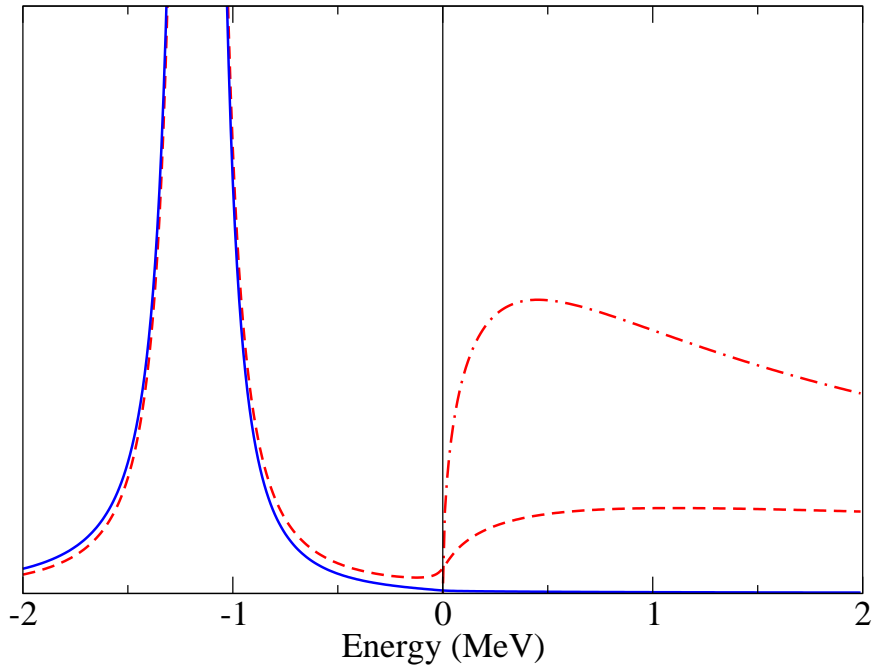}
\caption{
Line shapes of $X(3872)$ for 
$\gamma_{\rm re} + i \gamma_{\rm im} = 47.5$~MeV.
The curves are the line shape in $J/\psi \, \pi^+ \pi^-$ (solid line),
the line shape in $D^0 \bar D^0 \pi^0$ (dashed line), 
and the $D^{*0} \bar D^0$ energy distribution (dash-dotted line).
The two line shapes have been normalized so the resonances 
below the threshold have the same peak height.
\label{Fig:LSreal}}
\end{figure}
%%%%%%%%%%%%%%%%%%%%%%%%%%%%%%%%%%%%%%%%%%%%%%%%%%%%%%%%%%%%%%%%%%%%%%%%%%%%%%%%%%%%%%%%%

%%%%%%%%%%%%%%%%%%%%%%%%%%%%%%%%%%%%%%%%%%%%%%%%%%%%%%%%%%%%%%%%%%%%%%%%%%%%%%%%%%%%%%%%%
\begin{figure}[t]
\includegraphics[width=0.7\textwidth,clip=true]{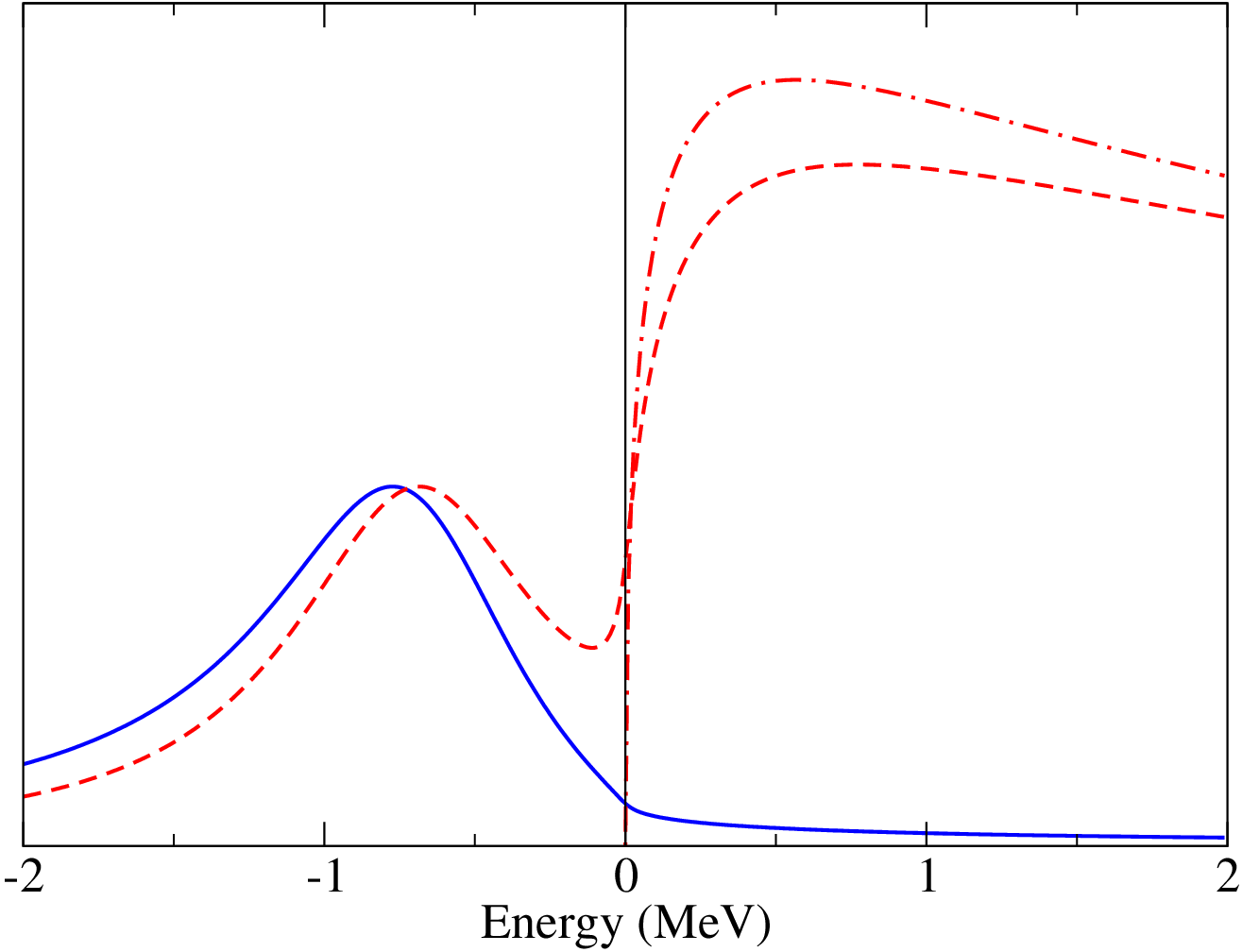}
\caption{
Line shapes of $X(3872)$ for 
$\gamma_{\rm re} + i \gamma_{\rm im} = (38.4 + 12.0 i)$~MeV.
The curves are the line shape in $J/\psi \, \pi^+ \pi^-$ (solid line),
the line shape in $D^0 \bar D^0 \pi^0$ (dashed line), 
and the $D^{*0} \bar D^0$ energy distribution (dash-dotted line).
The two line shapes have been normalized so the resonances 
below the threshold have the same peak height.
\label{Fig:LScomplex}}
\end{figure}
%%%%%%%%%%%%%%%%%%%%%%%%%%%%%%%%%%%%%%%%%%%%%%%%%%%%%%%%%%%%%%%%%%%%%%%%%%%%%%%%%%%%%%%%%

The fits to the Belle data give parameters $\gamma_{\rm re}$
and $\gamma_{\rm im}$ with smaller error bars than the fits to the
Babar data.  In Figs.~\ref{Fig:LSreal} and \ref{Fig:LScomplex}, 
the unsmeared line shapes of 
$X(3872)$ in the $J/\psi \, \pi^+ \pi^-$ decay channel
corresponding to the central values of the two fits 
to the Belle data are shown as solid lines.  
The line shape in Fig.~\ref{Fig:LScomplex} 
from using $\gamma_{\rm im}$ as a fitting parameter is wider 
than that in Figs.~\ref{Fig:LSreal} from setting $\gamma_{\rm im}=0$.
Both line shapes are much narrower than the smeared line shapes 
shown in Fig.~\ref{Fig:psipipiBelle}.
Thus most of the observed width 
can be accounted for by the experimental resolution.

\section{Energy distributions for the $\bm{D^0 \bar{D}^0 \pi^0}$ decay channel}
\label{sec:LS-long}

In this section, we summarize the essential aspects of the line shape 
of the $X(3872)$ in the $D^0 \bar D^0 \pi^0$ channel.
We also determine the energy distribution that follows from the
identification of $D^0 \bar D^0 \pi^0$ events with energy 
near the $D^{*0} \bar{D}^0$ threshold with
$D^{*0} \bar{D}^0$ and $D^{0} \bar{D}^{*0}$ events above the threshold.

%%%%%%%%%%%%%%%%%%%%%%%%%%%%%%%%%%%%%%%%%%%%%%%%%%%%%%%%%%%%%%%%%%%%%%%%%%%%%%%%%%%%%%%%%
\begin{figure}[t]
\includegraphics[width=0.8\textwidth,clip=true]{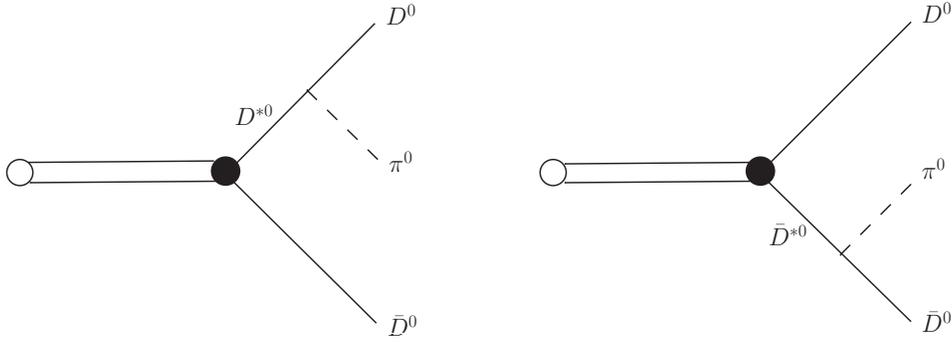}
\caption{
Diagrams for the production of $D^0 \bar D^0 \pi^0$.
The open dot represents the $B \to K$ transition that creates
$D^{*0} \bar D^0$ or $D^{0} \bar D^{*0}$ at a short-distance scale.  
The double line represents the propagation of the resonant linear 
combination of the pair of charm mesons.
The two diagrams involve either a virtual $D^{*0}$ (left diagram)
or a virtual $\bar D^{*0}$ (right diagram). 
\label{Fig:XDDpi}}
\end{figure}
%%%%%%%%%%%%%%%%%%%%%%%%%%%%%%%%%%%%%%%%%%%%%%%%%%%%%%%%%%%%%%%%%%%%%%%%%%%%%%%%%%%%%%%%%

In the decay $B^+ \to K^+ + D^0 \bar D^0 \pi^0$,
the momentum distributions for $D^0 \bar D^0 \pi^0$ 
near the $X(3872)$ resonance can be calculated 
from the sum of the two diagrams in Fig.~\ref{Fig:XDDpi}.  
The open dot represents the 
$B^+ \to K^+$ transition which creates a 
$D^{*0} \bar D^0$ or $D^{0} \bar D^{*0}$ at short distances.
The double line represents the exact propagator for the resonant 
superposition of $D^{*0} \bar D^0$ and $D^{0} \bar D^{*0}$,
whose dependence on the total energy $E$ of $D^0 \bar D^0 \pi^0$
is given by the scattering amplitude $f(E)$ in 
Eq.~(\ref{Eq:ScatteringAmplitude}).
In the propagators for the virtual $D^{*0}$ and $\bar D^{*0}$,
the width $\Gamma_{*0}$ must be taken into account.
The coupling of the $\pi^0$ to the charm mesons is linear in the
pion momentum.
The differential distribution in the total energy $E$ and in the momenta 
$\bm{p}_D$, $\bm{p}_{\bar  D}$, and $\bm{p}_\pi$ of the
$D^0$, $\bar D^0$, and $\pi^0$ has the form
\begin{equation}
d \Gamma \propto
|f(E)|^2 \, p_\pi^2 
\left| \frac{1}{p_D^2 - 2 \mu E - i \mu \Gamma_{*0}}
+ \frac{1}{p_{\bar  D}^2 - 2 \mu E - i \mu \Gamma_{*0}} \right|^2
d\Phi_{D \bar D \pi} \, dE ~.
\label{Eq:dGam}
\end{equation}
The differential 3-body phase space $d\Phi_{D \bar D \pi}$ includes a 
delta function that relates the energy $E$ and the three momenta:
\begin{equation}
E = - \delta_{D^*D\pi}
+ \frac{p_D^2}{2 M_{D^0}} + \frac{p_{\bar  D}^2}{2 M_{D^0}}
+ \frac{p_\pi^2}{2 m_{\pi^0}}~,
\label{Eq:econ}
\end{equation}
where $\delta_{D^*D\pi}$ is the energy released in the decay of 
$D^{*0}$ to $D^0 \pi^0$:
\begin{equation}
\delta_{D^*D\pi} \equiv M_{D^{*0}} - M_{D^0} - m_{\pi^0}
                        = 7.14 \pm 0.07~{\rm MeV}.
\label{Eq:deltaDstarDpi}
\end{equation}
The 3-body phase space can be reduced to a 2-dimensional integral over 
$p_D$ and $p_{\bar  D}$:
\begin{equation}
d\Phi_{D \bar D \pi} = 
\frac{2 m_{\pi^0}}{(2 \pi)^3} \, p_D d p_D \, p_{\bar  D} dp_{\bar  D}.
\label{dPhi2}
\end{equation}
The integration region is
\begin{equation}
\frac{p_D^2 + p_{\bar  D}^2}{2 \mu_{D \pi}} 
     - \frac{p_D p_{\bar  D}}{m_{\pi^0}}
< \delta_{D^*D\pi} + E <
\frac{p_D^2 + p_{\bar  D}^2}{2 \mu_{D \pi}} 
     + \frac{p_D p_{\bar  D}}{m_{\pi^0}} ,
\label{intpp}
\end{equation}
where $\mu_{D\pi}= M_{D^0} m_{\pi^0}/ (M_{D^0} + m_{\pi^0})$
is the reduced mass of $D^0$ and $\pi^0$.

Voloshin has used the diagrams in Fig.~\ref{Fig:XDDpi} to predict the 
momentum distributions for $D^0 \bar D^0 \pi^0$ in decays 
of the $X(3872)$ resonance \cite{Voloshin:2003nt}.  
His result is proportional to the right side of 
Eq.~(\ref{Eq:dGam}) with the resonance factor $|f(E)|^2$ omitted,
$\Gamma_{*0}$ set to 0, and the energy $E$ in the propagators 
replaced by $-E_X$, where $E_X$ is the binding energy of the $X(3872)$.
This is the appropriate momentum
distribution only if the energy $E$ is fixed at a value close to the peak 
of the resonance.  In the case of a low-energy antiproton beam
incident on a nucleon target, it may be possible to tune the 
center-of-mass energy to the peak of the resonance.  
However in the case of $B$ meson decays, the $X(3872)$ resonance 
is produced with a variable energy $E$. 
Since the experimental resolution in $E$ is
larger than the width of the resonance,
it is necessary to take the resonance factor $|f(E)|^2$ into account.

If $|E|$ is small compared to
$2 (m_{\pi^0}/M_{D^0}) \delta_{D^*D\pi} = 1.04$~MeV,
the phase space integral in Eq.~(\ref{Eq:dGam}) 
can be evaluated analytically.%
\footnote{It is also necessary for $\Gamma_{*0}/2$ to be small 
compared to $ (m_{\pi^0}/M_{D^0}) \delta_{D^*D\pi}$,
but this is satisfied if $\Gamma_{*0} \approx 65.5$~keV.}
In this case, the interference term between the two propagators in 
Eq.~(\ref{Eq:dGam}) can be neglected.   The 3-body phase space
integral in Eq.~(\ref{Eq:dGam}) reduces to
\begin{equation}
2 \int \frac{p_\pi^2}{|p_D^2 - 2 \mu E - i \mu \Gamma_{*0}|^2} 
d\Phi_{D \bar D \pi} \approx 
\frac{1}{\pi^2 \Gamma_{*0}} 
\left( \frac{2 \mu_{D\pi}^5 \delta_{D^*D\pi}^3}{\mu} \right)^{1/2}
\left( \sqrt{ E^2 + \Gamma_{*0}^2/4} + E \right)^{1/2} .
\label{int3}
\end{equation}
The resulting line shape has the form
\begin{equation}
\frac{d \Gamma}{dE} \propto
|f(E)|^2 \left( \sqrt{ E^2 + \Gamma_{*0}^2/4} + E \right)^{1/2}~.
\label{Eq:dGamdE}
\end{equation}
This simple expression for the line shape was first derived by 
Braaten and Lu \cite{Braaten:2007dw}.  If $\gamma_{\rm re}$ is positive, 
the line shape consists of a resonance 
associated with the bound state below the $D^{*0} \bar D^0$
threshold and a threshold enhancement above the threshold.
If $\gamma_{\rm re}$ is negative, there is 
a threshold enhancement above the $D^{*0} \bar D^0$ threshold
but no resonance below the threshold.
The position $E_{\rm max}^{D^{0} \bar D^0 \pi^0}$ 
of the maximum in the line shape satisfies
\begin{equation}
2 \mu \sqrt{E_{\rm max}^2 + \Gamma_{*0}^2/4} - 4 \mu E_{\rm max}
      + \gamma_{\rm re}^2 + \gamma_{\rm im}^2
- 4 \gamma_{\rm re}
\left( \mu \sqrt{E_{\rm max}^2 + \Gamma_{*0}^2/4} - \mu E_{\rm max} \right)^{1/2}
= 0~.
\label{Eq:eqEmaxDDpi}
\end{equation}
The solution up to corrections that are second order in $\Gamma_{*0}$ is
\begin{equation}
E_{\rm max}^{D^{0} \bar D^0 \pi^0} \approx 
- \frac{1}{2 \mu} 
\left( \frac{ 2 \gamma_{\rm re} + \sqrt{\gamma_{\rm re}^2 - 3 \gamma_{\rm im}^2}}
          {3} \right)^2 ~.
\label{Eq:EmaxDDpi}
\end{equation}

The normalization factor for the line shape in Eq.~(\ref{Eq:dGamdE}) 
involves the product $({\cal B} {\cal B})_{D^0 \bar D^0 \pi^0}$
of the branching fractions for $B^+ \to K^+ X$ and 
$X \to D^0 \bar D^0 \pi^0$ defined by Eq.~(\ref{Eq:BrBr}).
Defining these branching fractions is problematic, 
because the line shape in Eq.~(\ref{Eq:dGamdE}) is not integrable.
Since $|f(E)|^2$ decreases as $1/|E|$ for large $|E|$,
the integral of the line shape in Eq.~(\ref{Eq:dGamdE})
increases as the square root of the upper endpoint.
This implies that the product of branching fractions 
cannot be defined uniquely in terms of an integral over the line shape.
The numerical value of $({\cal B} {\cal B})_{D^0 \bar D^0 \pi^0}$ 
depends inevitably on the prescription used to define it.
Our prescription is that the normalized line shape 
for $B^\pm \to K^\pm + D^0 \bar D^0 \pi^0$ is
\begin{equation}
\frac{d\Gamma}{dE} \equiv
\Gamma[B^+] \, ({\cal B} {\cal B})_{D^0 \bar D^0 \pi^0} \,
\frac{d \hat \Gamma_{SD}}{dE}
\left( \frac{\sqrt{ E^2 + \Gamma_{*0}^2/4} + E}
       {\sqrt{ E_X^2 + \Gamma_{*0}^2/4} - E_X} \right)^{1/2}~,
\label{Eq:dGdEDDpi}
\end{equation}
where $d\hat \Gamma_{SD}/dE$ is the short-distance line shape
in Eq.~(\ref{Eq:dGhatdE})
and $E_X$ is the binding energy given by Eq.~(\ref{Eq:EX}). 
The last factor in Eq.~(\ref{Eq:dGdEDDpi}) reduces to 1 at $E = -E_X$.
In the case $\Gamma_X \ll 2 E_X$, the line shape in the region 
$|E + E_X| \ll E_X$ is approximately a Breit-Wigner resonance.
The integral of the right side of Eq.~(\ref{Eq:dGdEDDpi}) 
over this region is approximately 
$\Gamma[B^+] \, ({\cal B} {\cal B})_{D^0 \bar D^0 \pi^0}$,
justifying the interpretation of $({\cal B} {\cal B})_F$ as the 
product of the branching fraction for $B^+ \to K^+ + X$ and 
$X \to D^0 \bar D^0 \pi^0$.
If $\Gamma_X /(2 E_X)$ is not small, the constant 
$({\cal B} {\cal B})_{D^0 \bar D^0 \pi^0}$
defined by Eq.~(\ref{Eq:dGdEDDpi}) has no 
precise physical interpretation.  It is simply a convenient variable
for specifying the normalization of the line shape.

To compare with the energy distribution for $D^{*0} \bar D^0$
measured in the $B$ factory experiments, we must take into 
account how these energy distributions are measured.
Three particles identified as $D^0$, $\bar D^0$, and $\pi^0$ 
can be considered as candidates for either a $D^{*0} \bar D^0$
event or a $D^0 \bar D^{*0}$ event.  If the measured invariant mass of 
$D^0 \pi^0$ is close enough to the mass of $D^{*0}$
(within 10 MeV for Babar \cite{Babar:2007rva},
within 6 MeV for Belle \cite{Belle:2008su}), it is identified as a 
$D^{*0}$.  The constraint that the invariant mass of $D^0 \pi^0$ is 
equal to $M_{D^{*0}}$ is then used to sharpen the resolution of the
measured momenta.  If the $D^0$ and $\pi^0$ are produced by the decay 
of a constituent $D^{*0}$ from the bound state $X(3872)$, 
their invariant mass 
will be smaller than $M_{D^{*0}}$ by approximately the 
binding energy $E_X$.  This information about the binding energy 
is discarded when the $D^0 \pi^0$ is constrained to 
come from the decay of a $D^{*0}$.
If the momenta of the $D^0$, $\bar D^0$, and $\pi^0$ 
in the $D^0 \bar D^0 \pi^0$ rest frame are $\bm{p}_D$, 
$\bm{p}_{\bar  D}$, and $\bm{p}_\pi$, their total energy $E$
relative to the $D^{*0} \bar D^0$ threshold is
given in Eq.~(\ref{Eq:econ}).
If the $D^0 \pi^0$ is identified as a $D^{*0}$ in the experimental analysis,
the inferred energy $E_{\rm exp}$ of the $D^0 \bar D^0 \pi^0$ 
relative to the $D^{*0} \bar D^0$ threshold is
\begin{equation}
\frac{(\bm{p}_D + \bm{p}_\pi)^2}{2 M_{D^{*0}}} 
+ \frac{p_{\bar  D}^2}{2 M_{D^0}} 
= \frac{p_{\bar  D}^2}{2 \mu} ~.
\label{Eq:EDstarDbar}
\end{equation}
Similarly, if the $\bar D^0 \pi^0$ is identified as a $\bar D^{*0}$ 
in the experimental analysis, the inferred energy $E_{\rm exp}$ of the 
$D^0 \bar D^0 \pi^0$ relative to the $D^{*0} \bar D^0$ threshold is
$p_D^2/(2 \mu)$.  If the $D^0 \pi^0$ and $\bar D^0 \pi^0$ both
have invariant mass close enough to $M_{D^{*0}}$ to be 
identified as $D^{*0}$ and $\bar D^{*0}$, the one whose invariant 
mass is closest is constrained to be a $D^{*0}$ or $\bar D^{*0}$.
Thus the inferred energy $E_{\rm exp}$
of a $D^0 \bar D^0 \pi^0$ event that is identified as either 
$D^{*0} \bar D^0$ or $D^0 \bar D^{*0}$ is
\begin{subequations}
\begin{eqnarray}
E_{\rm exp} &=& 
\frac{{\rm min}(p_D^2,p_{\bar  D}^2)}{2 \mu} ~~~~~~
E < \frac{p_D^2 + p_{\bar  D}^2}{4 \mu},
\label{Eq:Eexp-less}
\\
&=& \frac{{\rm max}(p_D^2,p_{\bar  D}^2)}{2 \mu} ~~~~~~
E > \frac{p_D^2 + p_{\bar  D}^2}{4 \mu} ~.
\label{Eq:EDstarD}
\end{eqnarray}
\end{subequations}
We will refer to $E_{\rm exp}$ as the $D^{*0} \bar D^0$ energy.
It is the distribution in this variable 
that should be compared with the experimental energy distribution.

If $|E|$ is small compared to
$2 (m_{\pi^0}/M_{D^0}) \delta_{D^*D\pi} = 1.04$ MeV,
the line shape in the $D^0 \bar D^0 \pi^0$ channel is given 
by the analytic expression in Eq.~(\ref{Eq:dGdEDDpi}).
In this case it is also possible to obtain an analytic expression
for the distribution in the variable $E_{\rm exp}$.
For $|E| \ll 2 (m_{\pi^0}/M_{D^0}) \delta_{D^*D\pi}$,
the larger of the momenta $p_D$ and $p_{\bar  D}$ is 
approximately $(2 \mu_{D \pi} \delta_{D^*D\pi})^{1/2}$,
so $E_{\rm exp}$ is given by Eq.~(\ref{Eq:Eexp-less}).
The distribution in $E_{\rm exp}$ can then be expressed as 
\begin{eqnarray} 
\frac{d \Gamma}{d E_{\rm exp}} \approx
\frac{\Gamma[B^+] \, ({\cal B} {\cal B})_{D^0 \bar D^0 \pi^0} \, \Gamma_{*0}}
{\sqrt{2} \pi 
 \big( \sqrt{E_X^2 + \Gamma_{*0}^2/4} - E_X \big)^{1/2}}
E_{\rm exp}^{1/2}
\int_{-\infty}^\infty dE \, \frac{d \hat \Gamma_{SD}}{dE}
\frac{1}{|E_{\rm exp}- E - i \Gamma_{*0}/2|^2} ~,
\label{Eq:dGamdEDstarD}
\end{eqnarray}
where $\hat \Gamma_{SD}/dE$ is given in Eq.~(\ref{Eq:dGhatdE})
and $E_X$ is given in Eq.~(\ref{Eq:EX}).
The normalization is consistent with that in Eq.~(\ref{Eq:dGdEDDpi}),
as can be verified by integrating 
over $E_{\rm exp}$ using the integration formula
\begin{equation}
\int_0^\infty dE_{\rm exp} \, 
\frac{E_{\rm exp}^{1/2}}{|E_{\rm exp}- E - i \Gamma_{*0}/2|^2} = 
\frac{\sqrt{2} \pi}{\Gamma_{*0}}
\left( \sqrt{E^2 + \Gamma_{*0}^2/4} + E \right)^{1/2}~.
%\label{Eq:econ}
\end{equation}
The integral over $E$ in Eq.~(\ref{Eq:dGamdEDstarD})
can be evaluated analytically by deforming the integration contour
into the upper half-plane and picking up the contributions 
from the two poles and the branch cut.  The resulting expression 
for the integral over $E$ in Eq.~(\ref{Eq:dGamdEDstarD}) reduces to
\begin{eqnarray} 
\int_{-\infty}^\infty dE \, \frac{d \hat \Gamma_{SD}}{dE}
\frac{1}{|E_{\rm exp}- E - i \Gamma_{*0}/2|^2} &=&
\frac{\mu^2 \Gamma_X}{2 \pi \Gamma_{*0} |\gamma|^2}
\nonumber
\\
&& \hspace{-5cm}
\times \left( 
\frac{2 i \gamma_0^2 F(-i \gamma^2)}
    {(\gamma^2 - \gamma_*^2 + 2 i \gamma_0^2)
    (\gamma^2 + 2 \mu E_{\rm exp})
    (\gamma^2 + 2 \mu E_{\rm exp} + 2i \gamma_0^2)}
\right.
\nonumber
\\
&& \hspace{-4cm}
- \frac{2 i \gamma_0^2 \big[ F(-i \gamma_*^2 - 2 \gamma_0^2)
       - 2 \pi \big( \gamma_* + i \sqrt{-\gamma_*^2} \big)
       \big( \gamma + \sqrt{\gamma_*^2 - 2i \gamma_0^2} \big) \big]}
    {(\gamma^2 - \gamma_*^2 + 2 i \gamma_0^2)
    (\gamma_*^2 + 2 \mu E_{\rm exp})
    (\gamma_*^2 + 2 \mu E_{\rm exp} - 2i \gamma_0^2)}
\nonumber
\\
&& \hspace{-4cm} 
- \frac{F(2 i \mu E_{\rm exp} - 2 \gamma_0^2)
       - 2 \pi \big( \gamma_* + i \sqrt{2 \mu E_{\rm exp}} \big)
       \big( \gamma - i \sqrt{2 \mu E_{\rm exp} + 2i \gamma_0^2} \big)}
      {(\gamma_*^2 + 2 \mu E_{\rm exp})
       (\gamma^2 + 2 \mu E_{\rm exp} + 2i \gamma_0^2)}
\nonumber
\\
&& \hspace{-4cm} \left.
+ \frac{F(2 i \mu E_{\rm exp})}
      {(\gamma^2 + 2 \mu E_{\rm exp}) 
       (\gamma_*^2 + 2 \mu E_{\rm exp} - 2i \gamma_0^2)}
\right),
\label{Eq:intdGamdEdenIm}
\end{eqnarray}
where $\gamma_0 = (\mu \Gamma_{*0})^{1/2}$,
$\gamma = \gamma_{\rm re} + i \gamma_{\rm im}$,
$\gamma_* = \gamma_{\rm re} - i \gamma_{\rm im}$,
and the function $F(z)$ is
\begin{equation} 
F(z) = 
i \sqrt{-i(z + 2 \gamma_0^2)} 
\left( 2 \pi \gamma
 - 4 \sqrt{- i z}  \, 
\log \frac{(1+i)(\sqrt{- i z} + \sqrt{-i (z + 2 \gamma_0^2)}}
       {2 \gamma_0} \right).
\label{Eq:FdefIm}
\end{equation}
This function has a square-root branch point at $z = - 2 \gamma_0^2$,
but despite the factors of $\sqrt{z}$ it has no branch point at $z=0$.
Although it is not manifest, the expression on the right side of 
Eq.~(\ref{Eq:intdGamdEdenIm}) is real-valued.  

In Figs.~\ref{Fig:LSreal} and \ref{Fig:LScomplex}, the solid lines are
the line shapes in the $J/\psi \, \pi^+ \pi^-$ decay channel for
$\gamma_{\rm re} + i \gamma_{\rm im} = 47.5$~MeV 
and $(38.4 + 12.0 i)$~MeV, respectively.  For comparison,
the line shapes in the $D^0 \bar D^0 \pi^0$ decay channel 
and the $D^{*0} \bar D^0$ energy distributions 
are also shown as dashed and dash-dotted lines, respectively. 
In each figure, the curves are normalized so that the resonances 
below the threshold have the same maximum values.
In both figures, the $D^0 \bar D^0 \pi^0$ line shape has a peak below the
$D^{*0} \bar D^0$ threshold corresponding to the $X(3872)$ resonance
and a second peak above the threshold corresponding to a
threshold enhancement in the production of $D^{*0} \bar D^0$ 
and $D^{0} \bar D^{*0}$.  The position and width of the resonance peak 
is close to that for the $J/\psi \, \pi^+ \pi^-$ line shape. 
The $D^{*0} \bar D^0$ energy distribution, which vanishes below the 
threshold, has a peak above the threshold whose width 
is considerably larger than the width of the resonance.
Thus a measurement of the position and width of the peak 
in the $D^{*0} \bar D^0$ invariant mass distribution should not be 
interpreted as a measurement of the
mass and width of the $X(3872)$.

\section{Analysis of the $\bm{D^0 \bar{D}^0 \pi^0}$ decay channel}
\label{sec:DDpi}

%%%%%%%%%%%%%%%%%%%%%%%%%%%%%%%%%%%%%%%%%%%%%%%%%%%%%%%%%%%%%%%%%%%%%%%%%%%%%%%%%%%%%%%%%
\begin{figure}[t]
\includegraphics[width=0.8\textwidth,clip=true]{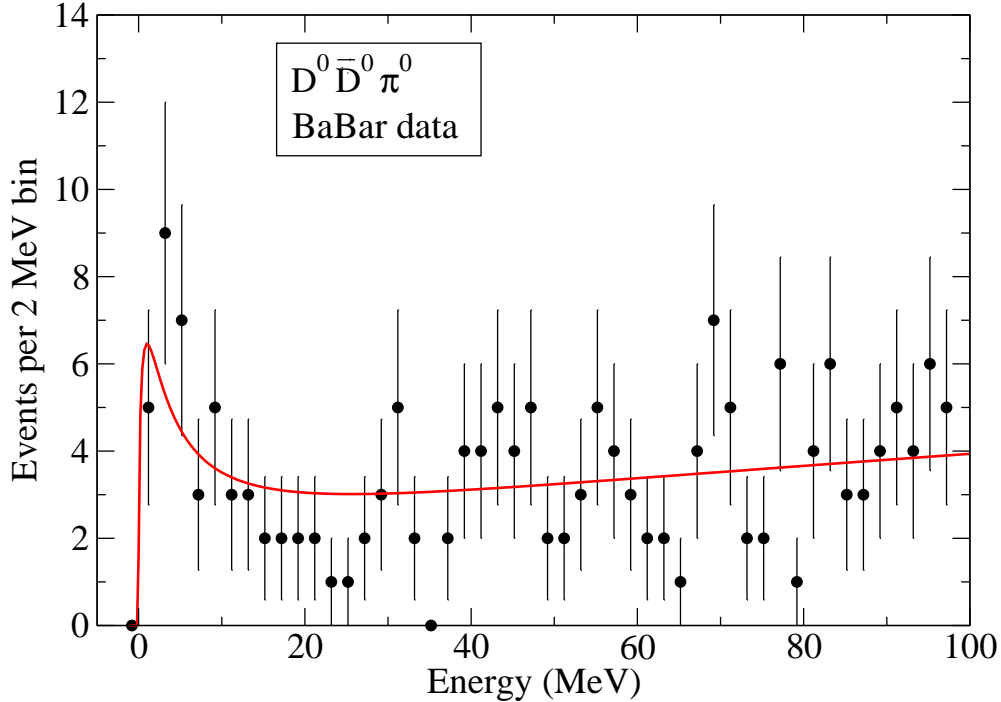}
\caption{
Energy distribution for $D^{*0} \bar D^0$ and $D^{0} \bar D^{*0}$ 
measured by the Babar Collaboration \cite{Babar:2007rva}.
The data are the number of events per 2~MeV bin.
The inverse scattering length $\gamma_{\rm re} + i \gamma_{\rm im}$
for the fit is 67.7~MeV.
\label{Fig:DDpiBabar}}
\end{figure}
%%%%%%%%%%%%%%%%%%%%%%%%%%%%%%%%%%%%%%%%%%%%%%%%%%%%%%%%%%%%%%%%%%%%%%%%%%%%%%%%%%%%%%%%%

%%%%%%%%%%%%%%%%%%%%%%%%%%%%%%%%%%%%%%%%%%%%%%%%%%%%%%%%%%%%%%%%%%%%%%%%%%%%%%%%%%%%%%%%%
\begin{figure}[t]
\includegraphics[width=0.8\textwidth,clip=true]{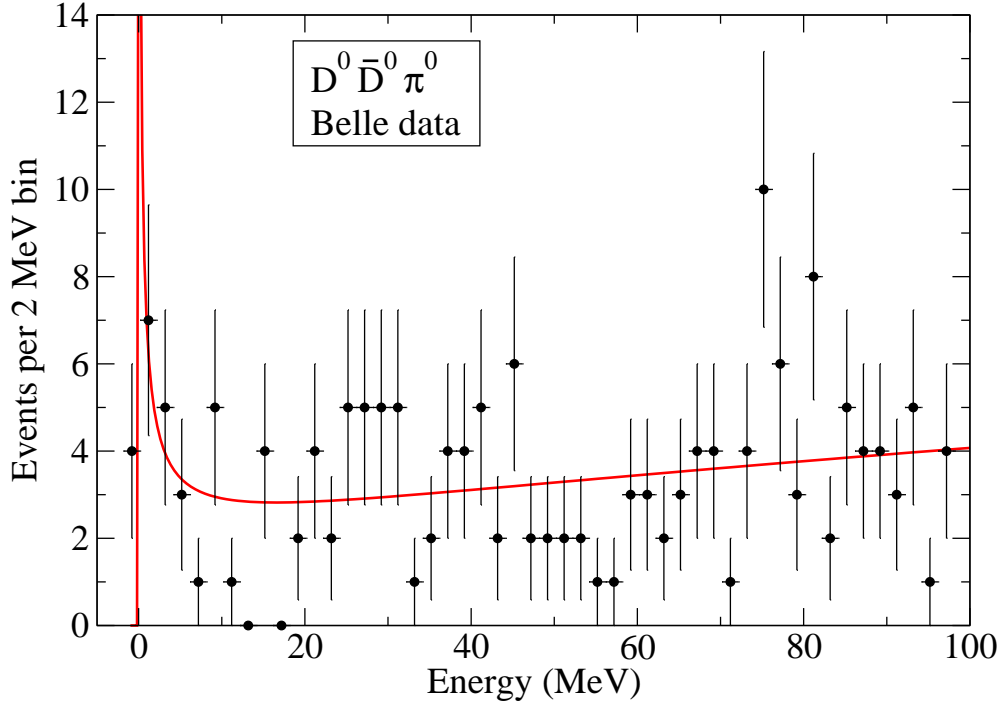}
\caption{
Energy distribution for $D^{*0} \bar D^0$ and $D^{0} \bar D^{*0}$ 
measured by the Belle Collaboration \cite{Belle:2008su}.
The data are the number of events per 2~MeV bin.
The inverse scattering length $\gamma_{\rm re} + i \gamma_{\rm im}$
for the fit is 9.99~MeV.
\label{Fig:DDpiBelle}}
\end{figure}
%%%%%%%%%%%%%%%%%%%%%%%%%%%%%%%%%%%%%%%%%%%%%%%%%%%%%%%%%%%%%%%%%%%%%%%%%%%%%%%%%%%%%%%%%

In this section, we analyze recent data from the Belle and Babar 
Collaborations on the line shape of the $X(3872)$ in the 
$D^0 \bar D^0 \pi^0$ decay mode \cite{Babar:2007rva,Belle:2008su}.
We consider the energy distribution for $D^{*0} \bar D^0$ 
and $D^{0} \bar D^{*0}$ in the interval from 0 to 100~MeV.
For our two data samples, the total number $N_{B \bar B}$ of $B^+ B^-$ 
and $B^0 \bar B^0$ events accumulated
and the number of candidate events for the decay of $B^\pm$ into 
$K^\pm + D^{*0} \bar D^0 (D^{0} \bar D^{*0})$ are as follows:
\begin{itemize}
\item
Babar Collaboration \cite{Babar:2007rva}:
$N_{B \bar B} = 3.83 \times 10^8$, 
172 events in 50 bins of width 2 MeV,
\item
Belle Collaboration \cite{Belle:2008su}:
$N_{B \bar B} = 6.57 \times 10^8$, 
171 events in 50 bins of width 2 MeV.
\end{itemize}
The data are shown in Figs.~\ref{Fig:DDpiBabar} and \ref{Fig:DDpiBelle}.
The vertical error bar in a bin with $n$ events is $\sqrt{n}$.
The horizontal error bar indicates the width of the bin.

We take the distribution in the $D^{*0} \bar D^0$ energy $E_{\rm exp}$
to be given by Eqs.~(\ref{Eq:dGamdEDstarD}),
(\ref{Eq:intdGamdEdenIm}), and (\ref{Eq:FdefIm}).
This energy distribution should be accurate within about an MeV 
of the threshold.  We assume that the dominant contributions 
to the signal come from this threshold region.
To predict the number of events in a given bin of invariant mass,
we need to take into account the background and the energy resolution 
of the experiment.  
The predicted number of $D^{*0} \bar D^0$ events
in an energy bin of width $\Delta$ centered at $E_i$ 
can be expressed as
\begin{eqnarray}
N_i &=& 2 N_{B \bar B}  \, 
\int_{E_i - \Delta/2}^{E_i + \Delta/2} dE'
\int_{0}^\infty dE_{\rm exp} \, R(E',E_{\rm exp})  \, E_{\rm exp}^{1/2}
\nonumber
\\
&& \times 
\left[ 
\frac{({\cal B} {\cal B})_{D^0 \bar D^0 \pi^0} \Gamma_{*0}}
    {\sqrt{2} \pi 
     \big( \sqrt{E_X^2 + \Gamma_{*0}^2/4} - E_X \big)^{1/2}}
\int_{-\infty}^\infty dE \, \frac{d \hat \Gamma_{SD}}{dE} 
       \frac{1}{|E_{\rm exp}- E - i \Gamma_{*0}/2|^2}
     + C_{\rm bg} \right]~,
\nonumber
\\
\label{Eq:N-DDpi}
\end{eqnarray}
where $C_{\rm bg}$ takes into account the background. 
Our energy interval 0--100 MeV is narrow enough 
that the background contribution to the distribution in $E_{\rm exp}$
can be taken as a constant 
$C_{\rm bg}$ multiplied by $E_{\rm exp}^{1/2}$, 
which is the energy dependence of the
$D^{*0} \bar D^0$ phase space.
The experimental resolution is taken into account through
a convolution with a Gaussian resolution function
with an energy-dependent width $\sigma(E_{\rm exp})$:
\begin{equation}
R(E',E_{\rm exp}) =
\frac{1}{\sqrt{2 \pi} \sigma(E_{\rm exp})} \,
\exp(-(E'-E_{\rm exp})^2/(2\sigma(E_{\rm exp})^2)) ~.
\label{Eq:RE'}
\end{equation}
We follow Ref.~\cite{Zhang:2009bv} in taking the width for both experiments 
to be the same energy-dependent function: 
\begin{equation}
\sigma(E_{\rm exp}) = \sqrt{(0.031~{\rm MeV})E_{\rm exp}}~.
\label{smear:DDpi}
\end{equation}
This may be too crude a model for the effects of the experimental 
resolution in this channel, 
but we will use it for illustrative purposes anyway.

%------------------------------------------------------------------------------------
\begin{table}[t]
\begin{tabular}{l|ccc|cccc}
data set       & ~$\gamma_{\rm re}$~ & ~$\gamma_{\rm im}$~ & 
                 $({\cal B} {\cal B})_{D^0 \bar D^0 \pi^0}$& 
                 ~$-E_X$~ & ~$\Gamma_X$~ & $E_{\rm max}$ & $\Gamma_{\rm fwhm}$ \\
\hline
Babar &
$67.7^{+10.9}_{-\ 9.3}$ &         0     & $0.034^{+0.008}_{-0.007}$ & 
$-2.37^{+0.61}_{-0.82}$ & $0.066 \pm 0.015$ &
$-2.37^{+0.61}_{-0.82}$ & $0.066 \pm 0.015$ \\
Babar &
$67.7^{+12.3}_{-\ 9.3}$ & $0^{+0.14}_{-0}$  & $0.034^{+0.008}_{-0.007}$ & 
$-2.37^{+0.61}_{-0.94}$ & $0.066^{+0.025}_{-0.015}$ &
$-2.37^{+0.61}_{-0.94}$ & $0.066^{+0.025}_{-0.015}$ \\
\hline
Belle &
$9.99^{+1.99}_{-1.42}$ &         0     & $0.0029^{+0.0007}_{-0.0006}$ & 
$-0.052^{+0.014}_{-0.023}$ & $0.066 \pm 0.015$ &
$-0.056^{+0.013}_{-0.022}$ & $0.066 \pm 0.015$ \\
Belle &
$9.99^{+3.16}_{-1.42}$ & $0^{+0.98}_{-0}$  & $0.0029^{+0.0007}_{-0.0006}$ & 
$-0.052^{+0.014}_{-0.038}$ & $0.066^{+0.025}_{-0.015}$ &
$-0.056^{+0.013}_{-0.037}$ & $0.066^{+0.026}_{-0.015}$ 
\end{tabular}
\caption{Results of our analyses of the data for 
$B^\pm \to K^\pm + D^0 \bar D^0 \pi^0$.
The four rows correspond to analyses using either 
the Belle data \cite{Belle:2008su} or the Babar data \cite{Babar:2007rva}
and either setting $\gamma_{\rm im} = 0$ 
or using $\gamma_{\rm im}$ as a fitting parameter.  
All entries are in units of MeV, except for 
$({\cal B} {\cal B})_{D^0 \bar D^0 \pi^0}$, 
which is in units of $10^{-6}$.}
\label{tab:DDpi}
\end{table}

We assume that the number of events in each bin of the smeared energy
$E'$ has a Poisson distribution whose mean 
value is given by $N_i$ in Eq.~(\ref{Eq:N-DDpi}).
We fix the $D^{*0}$ width $\Gamma_{*0}$ at 65.5~keV.
The adjustable parameters are $\gamma_{\rm re}$, $\gamma_{\rm im}$,
$({\cal B} {\cal B})_{D^0 \bar D^0 \pi^0}$, and $C_{\rm bg}$.
We determine the best fit to these parameters by maximizing the
likelihood for the observed distribution.  
For both the Belle and Babar data sets, 
we carry out two fits:  one with $\gamma_{\rm im} = 0$ and one with
$\gamma_{\rm im}$ as a fitting parameter.  
The results of our four analyses are presented in Table~\ref{tab:DDpi}.
The error bars
are determined in the same way as those in Table~\ref{tab:psipipi},
except that the uncertainty of $\pm 0.36$~MeV in the $D^{*0} \bar D^0$ 
threshold energy does not enter because the experimental energies 
were measured relative to this threshold.

%%%%%%%%%%%%%%%%%%%%%%%%%%%%%%%%%%%%%%%%%%%%%%%%%%%%%%%%%%%%%%%%%%%%%%%%%%%%%%%%%%%%%%%%%
\begin{figure}[t]
\includegraphics[width=0.7\textwidth,clip=true]{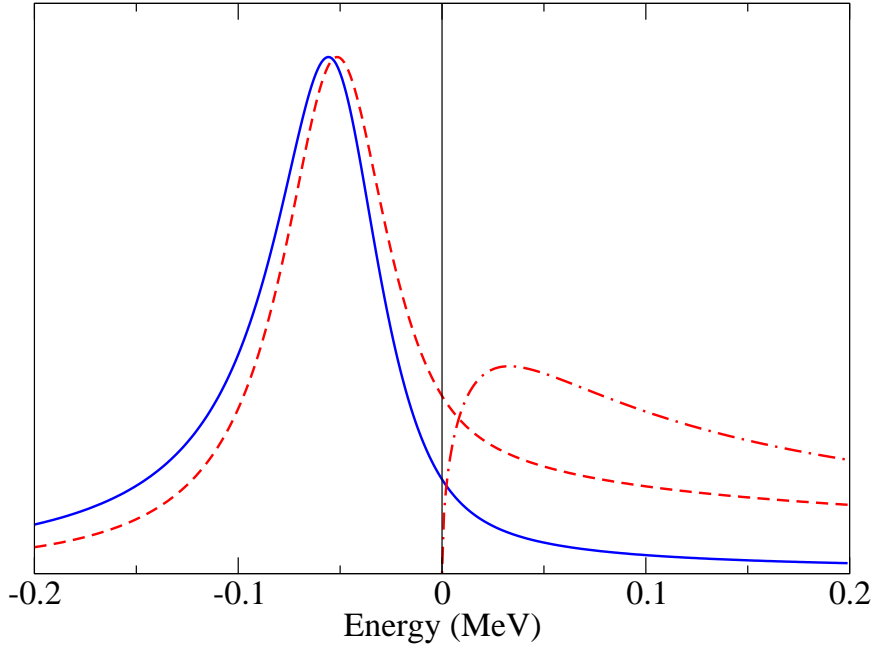}
\caption{
Line shapes of $X(3872)$ for 
$\gamma_{\rm re} + i \gamma_{\rm im} = 9.99$~MeV.
The curves are the line shape in $J/\psi \, \pi^+ \pi^-$ (solid line),
the line shape in $D^0 \bar D^0 \pi^0$ (dashed line), 
and the $D^{*0} \bar D^0$ energy distribution (dash-dotted line).
The two line shapes have been normalized so the resonances 
below the threshold have the same peak height.
\label{Fig:LSfit}}
\end{figure}
%%%%%%%%%%%%%%%%%%%%%%%%%%%%%%%%%%%%%%%%%%%%%%%%%%%%%%%%%%%%%%%%%%%%%%%%%%%%%%%%%%%%%%%%%

In the fits to the Babar and Belle data with $\gamma_{\rm im}$ 
treated as a fitting parameter, the maximum likelihood 
is obtained for $\gamma_{\rm im} = 0$, which is the 
smallest possible physical value.  This suggests that our model
for the experimental resolution in Eq.~(\ref{smear:DDpi}) may 
provide too much smearing of the energy distribution.
The best fit to the Babar data gives a line shape whose peak 
is below the $D^{*0} \bar D^0$ threshold by about 2.4~MeV, which is 
incompatible with the assumption $|E| \ll 1$~MeV that we used 
to derive analytic expressions for the line shape and the energy 
distribution.  The best fit to the Belle data gives a line shape 
whose peak is below the $D^{*0} \bar D^0$ threshold by only about
0.05~MeV, which is comparable to the width $\Gamma_X \approx 0.07$~MeV.  
The value of $({\cal B} {\cal B})_{D^0 \bar D^0 \pi^0}$
from the fit to the Belle data is about an order of magnitude smaller 
than that from the fit to the Babar data.
Since the Belle fit does not satisfy $\Gamma_X \ll 2 E_X$,
the value of $({\cal B} {\cal B})_{D^0 \bar D^0 \pi^0}$ should not 
be interpreted literally as the product of branching fractions.
It is simply a parameter used to specify the normalization 
of the line shape in Eq.~(\ref{Eq:dGdEDDpi}).  
The Babar fit does satisfy $\Gamma_X \ll 2 E_X$, so the value 
of $({\cal B} {\cal B})_{D^0 \bar D^0 \pi^0}$ can be 
interpreted as the product of branching fractions.
Dividing by the value $8.7 \times 10^{-6}$ 
for $({\cal B} {\cal B})_{J/\psi \, \pi^+ \pi^-}$ from 
Table~\ref{tab:psipipi}, we obtain a branching ratio for  
$D^0 \bar D^0 \pi^0$ to $J/\psi \, \pi^+ \pi^-$ of approximately 0.004.
This implies that short-distance decay modes 
account for most of the width $\Gamma_X$ of the $X(3872)$
resonance.  However the value $\gamma_{\rm im} = 0$ for the best fit
implies that the contribution to the width $\Gamma_X$ 
from short-distance decay modes is negligible.
A possible explanation for this inconsistency is that the
simple model for the $D^{*0} \bar D^0$ energy resolution
given in Eqs.~(\ref{Eq:RE'}) and (\ref{smear:DDpi}) is inadequate.

In Fig.~\ref{Fig:LSfit}, we show the line shapes corresponding 
to the best fit to the Belle data for $D^0 \bar D^0 \pi^0$.
The line shape in the $J/\psi \, \pi^+ \pi^-$ decay channel,
the line shape in the $D^0 \bar D^0 \pi^0$ decay channel, 
and the $D^{*0} \bar D^0$ energy distribution
are shown as solid, dashed, and dash-dotted lines, respectively.
The curves are normalized so that the resonances below the 
threshold have the same peak height.
The $D^0 \bar D^0 \pi^0$ line shape has a single peak below the
$D^{*0} \bar D^0$ threshold whose position and width are close 
to those for the peak in the $J/\psi \, \pi^+ \pi^-$ line shape.
It is this peak that should be identified with the $X(3872)$ resonance.
The $D^{*0} \bar D^0$ energy distribution, which vanishes below 
the $D^{*0} \bar D^0$ threshold, has a peak above the threshold
whose width is considerably larger than that of the $X(3872)$ 
resonance.  It is also much narrower 
than the smeared energy distribution shown in Fig.~\ref{Fig:DDpiBelle}.

\section{Critique of Previous Analyses}
\label{sec:critique}

In this section, we discuss how the analysis presented in this paper 
could be improved.  We also point out errors and misconceptions 
in previous theoretical analyses of the line shapes of the $X(3872)$.

The most limiting aspect of our analysis was the use of the 
analytic expression in Eq.~(\ref{Eq:dGdEDDpi})
for the line shape in the $D^0 \bar D^0 \pi^0$ decay channel.
The derivation of this expression involved the assumption
$|E| \ll 2 (m_{\pi^0}/M_{D^0}) \delta_{D^*D\pi} = 1.04$~MeV.
It requires most of the  $D^0 \bar D^0 \pi^0$ events
to be produced with energy within an MeV of the 
$D^{*0} \bar D^0$ threshold.  In particular, it requires the width
of the $X(3872)$ resonance to be much less than an MeV.
This limitation can be relaxed by replacing the 
invariant mass distribution in Eq.~(\ref{Eq:dGdEDDpi})
by the differential decay distribution in Eq.~(\ref{Eq:dGam}). 
Given our prescription for $({\cal B} {\cal B})_{D^0 \bar D^0 \pi^0}$
in Eq.~(\ref{Eq:dGdEDDpi}), the normalization 
of the differential decay rate is determined:
\begin{eqnarray}
\frac{d \Gamma}{dE} &=&
\frac{\Gamma[B^+] ({\cal B} {\cal B})_{D^0 \bar D^0 \pi^0} \, \pi^2 \Gamma_{*0}}
{(\sqrt{ E_X^2 + \Gamma_{*0}^2/4} + E_X)^{1/2}}
\left( \frac{\mu}{2 \mu_{D \pi}^5 \delta_{D^*D\pi}^3} \right)^{1/2}
\nonumber
\\
&& \times \frac{d \hat \Gamma_{SD}}{dE} \, p_\pi^2 
\left| \frac{1}{p_D^2 - 2 \mu E - i \mu \Gamma_{*0}}
+ \frac{1}{p_{\bar  D}^2 - 2 \mu E - i \mu \Gamma_{*0}} \right|^2
d\Phi_{D \bar D \pi} ~,
\label{Eq:dGamDDpi}
\end{eqnarray}
where $d\Phi_{D \bar D \pi}$ is given in Eq.~(\ref{dPhi2}).
The integral over the momenta $p_D$ and $p_{\bar D}$
must be evaluated numerically.

Another limiting aspect of our analysis was ignoring the effects of 
the charged charm meson pairs $D^{*+} D^-$ and $D^{+} D^{*-}$. 
They can produce significant interference effects for $|E|$ 
as small as 1/4 of the 8.1~MeV splitting between the $D^{*+} D^-$
and $D^{*0} \bar D^0$ thresholds \cite{Braaten:2007ft}.
The effects of charged charm meson pairs were first considered 
by Voloshin \cite{Voloshin:2007hh}, 
but there were conceptual errors in his analysis.
A correct analysis was presented by Braaten and Lu \cite{Braaten:2007ft}.
It involves the $2 \times 2$ matrix of S-wave $C=+$
scattering amplitudes $f_{ij}(E)$ between the neutral channel
$D^{*0} \bar D^0/D^{0} \bar D^{*0}$ labelled by subscript 0
and the charged channel
$D^{*+} D^-/D^{+} D^{*-}$ labelled by subscript 1.
The parameters in these scattering amplitudes are the inverse scattering
lengths $\gamma_0$ and $\gamma_1$ for charm mesons in the channels 
with isospin 0 and 1 in the isospin symmetry limit.
The coupled-channel expressions 
for the $D^{*} \bar D$ scattering amplitudes are 
%-----------------
\begin{subequations}
\begin{eqnarray}
f_{00}(E) &=&
\frac{- (\gamma_0 + \gamma_1) + 2 \kappa_1(E)}{D(E)} ,
\label{f00-E}
\\
f_{01}(E) &=& f_{10}(E) =
\frac{\gamma_1 - \gamma_0}{D(E)} ,
\label{f01-E}
\\
f_{11}(E) &=&
\frac{- (\gamma_0 + \gamma_1) + 2 \kappa(E)}{D(E)} ,
\label{f11-E}
\end{eqnarray}
\label{fij-E}
\end{subequations}
%-----------------
where the denominator is 
%-----------------
\begin{equation}
D(E) = 2 \gamma_0 \gamma_1 
- (\gamma_0 + \gamma_1) [\kappa(E) +  \kappa_1(E)]
+ 2 \kappa_1(E) \kappa(E) .
\label{D-E}
\end{equation}
%-----------------
The functions $\kappa(E)$ and $\kappa_1(E)$ are
%-----------------
\begin{subequations}
\begin{eqnarray}
\kappa(E) &=& \sqrt{- 2 \mu [E + i \Gamma_{*0}/2]} ,
\label{kappa-E}
\\
\kappa_1(E) &=&
\sqrt{- 2 \mu_1 [E - \nu + i \Gamma_{*1}/2]} ,
\label{kappa1-E}
\end{eqnarray}
\end{subequations}
%-----------------
where $\mu_1= 968.7$~MeV is the reduced mass of $D^{*+}$ and $D^-$ 
and $\nu = 8.1$~MeV is the splitting between the $D^{*+} D^-$
and $D^{*0} \bar D^0$ thresholds.
In Eq.~(\ref{kappa-E}), $\Gamma_{*0}$ is the energy-dependent width 
of the $D^{*0}$, which has its physical value 65.5~keV at $E=0$.
In Eq.~(\ref{kappa1-E}), $\Gamma_{*1}$ is the energy-dependent width 
of the $D^{*+}$, which has its physical value 96~keV 
at $E = 8.1$~MeV and decreases to 1.5~keV at $E= 2.1$~MeV,
which is the $D^+ D^- \pi^0$ threshold.
Near the $D^{*0} \bar D^0$ threshold, the scattering 
amplitude in Eq.~(\ref{f00-E}) reduces to the universal expression 
in Eq.~(\ref{Eq:ScatteringAmplitude}) with
%-----------------
\begin{equation}
\gamma_{\rm re} + i \gamma_{\rm im}  =
\frac{2 \gamma_0 \gamma_1 - (\gamma_0 + \gamma_1) \kappa_1(0)}
    {(\gamma_0 + \gamma_1) - 2 \kappa_10)} .
\label{gamma-CC}
\end{equation}
%-----------------

Voloshin's first conceptual error in Ref.~\cite{Voloshin:2007hh} 
was assuming that a $B \to K$ 
transition that produces the $X(3872)$ resonance must create the
charm mesons in the neutral channel and not in the charged channel.
This implies that the amplitudes for the resonant production of an 
isospin-0 final state such as $J/\psi \, \pi^+ \pi^- \pi^0$ and an 
isospin-1 final state such as $J/\psi \, \pi^+ \pi^-$
are proportional to  $f_{00}(E) - f_{01}(E)$ and 
$f_{00}(E) + f_{01}(E)$, respectively.
However, since there is resonant scattering between the 
neutral and charged channels, the $X(3872)$ resonance can also be 
produced by a $B \to K$ transition that creates the
charm mesons in the charged channel.
Thus the amplitudes for isospin-0 and isospin-1 final states
can also have terms proportional to  $f_{10}(E) - f_{11}(E)$ and 
$f_{10}(E) + f_{11}(E)$, respectively.
Voloshin's second conceptual error was ignoring the constraints of
isospin symmetry on the amplitudes for the creation of charm meson 
pairs by the $B^+ \to K^+$ and $B^0 \to K^0$ transitions.
He concluded incorrectly that the rates are proportional to 
$|f_{00}(E) \pm f_{01}(E)|^2$ with the same multiplicative constant for 
$B^+$ and $B^0$ decays.  The isospin symmetry constraints 
derived in Ref.~\cite{Braaten:2007ft} imply 
that the line shapes are different for $B^+$ and $B^0$ decays
and they are determined by three independent multiplicative constants.

Another limiting aspect of our analysis was ignoring the effects of 
the P-wave charmonium state $\chi_{c1}(2P)$. 
This state has the same quantum numbers $1^{++}$ as the $X(3872)$,
so it also has an S-wave coupling to charm meson pairs
$D^* \bar D$ and $D \bar D^*$.
Its spin symmetry partner  $\chi_{c2}(2P)$, which was discovered 
in 2006 by the Belle Collaboration \cite{Uehara:2005qd}, 
has a mass of about 3930~MeV.  Potential models predict the mass 
of the $\chi_{c1}(2P)$ to be lower by 20 to 50 MeV
\cite{Barnes:2005pb,Eichten:2005ga}.
Thus its mass could be close enough to the $D^* \bar D$ thresholds 
for the resonant coupling of the $\chi_{c1}(2P)$ to the charm mesons to
affect the line shape in this region.  Its effects on the line shapes 
within about an MeV of the threshold would however be negligible.
To be more precise, these effects are accurately taken into account
through the value of the inverse scattering length
$\gamma_{\rm re}+ i \gamma_{\rm im}$.
This follows from the universal behavior of an S-wave threshold resonance
which makes it insensitive to the mechanism for the resonance.
If the mass of the $\chi_{c1}(2P)$ is extremely close to the threshold,
it is transformed by its resonant couplings to the charm mesons 
into a charm meson molecule whose constituents have a large mean 
separation.
Thus far there has been no quantitative analysis of the effects 
of the $\chi_{c1}(2P)$ on the line shape of the $X(3872)$. 

In Ref.~\cite{Hanhart:2007yq}, Hanhart, Kalashnikova, Kudryavtsev, 
and Nefediev analyzed 
the line shapes for the $X(3872)$ using a generalization 
of a parametrization of the line shape for a near-threshold resonance 
proposed by Flatt\'e \cite{Flatte:1976xu}. 
Their expression for the line shape can be written as $|f_{\rm HKKN}(E)|^2$,
where $f_{\rm HKKN}(E)$ is the scattering amplitude
%-----------------
\begin{eqnarray}
f_{\rm HKKN}(E) &=&
\frac{1}{(2/g_1)E_f - i \Gamma(E)/g_1 + \kappa(E) 
        + (g_2/g_1) \kappa_1(E) - (2/g_1)E} ,
\label{f-Flatte}
\end{eqnarray}
%-----------------
$\kappa(E)$ is given by Eq.~(\ref{kappa-E}) with $\Gamma_{*0} = 0$,
$\kappa_1(E)$ is given by Eq.~(\ref{kappa1-E}) with $\Gamma_{*1} = 0$, 
and $\Gamma(E)$ is the energy-dependent partial width for 
short-distance decays of the $X(3872)$.
This lineshape was also used in a recent analysis by 
Zhang, Meng and Zheng \cite{Zhang:2009bv}.
Near the $D^{*0} \bar D^0$ threshold, the scattering 
amplitude in Eq.~(\ref{f00-E}) reduces to the universal expression 
in Eq.~(\ref{Eq:ScatteringAmplitude}) with
$\gamma_{\rm re} = -[2 E_f + g_2 \kappa(0)]/g_1$ and 
$\gamma_{\rm im} = \Gamma(0)/g_1$.
The coefficient $-2/g_1$ of the term $- (2/g_1)E$ in the denominator
of Eq.~(\ref{f-Flatte}) can be identified as 
$r_s/2$, where $r_s$ is the effective range.
In Ref.~\cite{Hanhart:2007yq}, Hanhart et al.\ 
found that the best fits to the Belle and Babar data
are in a scaling region of the parameter space in which
the $(2/g_1)E$ term in the denominator is negligible.
Thus the only relevant parameters are $E_f/g_1$, $g_2/g_1$,
and $\Gamma(0)/g_1$.
The scattering amplitude in Eq.~(\ref{f-Flatte})
with the $(2/g_1)E$ term omitted can be obtained from
the coupled-channel scattering amplitude $f_{00}(E)$ in Eq.~(\ref{f00-E}) 
by replacing $\kappa_1(E)$ in both the numerator
and the last term in the denominator by $\kappa_1(0)$.
Thus the line shape $|f_{\rm HKKN}(E)|^2$ takes into account 
some of the effects of the coupling between the neutral 
and charged channels.  

One apparent advantage of the line shape $|f_{\rm HKKN}(E)|^2$
that is actually illusory is that it is an integrable function 
of the energy $E$.
The product of branching fractions $({\cal B} {\cal B})_F$
for a short-distance channel $F$ of $X(3872)$ in the decay 
$B^+ \to K^+ + X$
can therefore be defined in the conventional way by specifying the 
energy distribution for the final state $F$ to be
%-----------------
\begin{equation}
\frac{d \Gamma}{dE} = \Gamma[B^+] ({\cal B} {\cal B})_F 
|f_{\rm HKKN}(E)|^2 \Big/ \int^\infty_{-\infty} dE'\, |f_{\rm HKKN}(E')|^2 .
\label{fHKKN-psipipi}
\end{equation}
%-----------------
This definition of $({\cal B} {\cal B})_F$ appears to be more natural
than the prescription for an threshold 
S-wave resonance that we introduced in Eq.~(\ref{Eq:dGdE}).
However the term in Eq.~(\ref{f-Flatte}) that makes the line shape 
$|f_{\rm HKKN}(E)|^2$ integrable is the last term $- (2/g_1)E$.  
Because the best fit is in a scaling region of the parameter space,
varying the parameter $g_1$ while holding the combinations 
$E_f/g_1$, $g_2/g_1$, and $\Gamma(0)/g_1$ fixed has essentially 
no effect on the line shape very near the resonance
but it does change the interval of the energy $E$ that gives 
significant contributions to the integral of the line shape.  
The numerical value of $({\cal B} {\cal B})_F$ is therefore determined 
by the value of $g_1$.  Thus the definition for 
$({\cal B} {\cal B})_F$ in Eq.~(\ref{f-Flatte}) is in fact 
an arbitrary prescription 
specified by the parameter $g_1$.

In their analysis of the line shape of $X(3872)$ in the
$D^0 \bar D^0 \pi^0$ decay mode in Ref.~\cite{Hanhart:2007yq}, 
Hanhart et al.\  made a serious conceptual error.
They assumed that the line shape has the form
%-----------------
\begin{equation}
\frac{d \Gamma}{dE} \propto
|f_{\rm HKKN}(E)|^2 \, E^{1/2} \, \theta(E).
\label{fHKKN-DDpi}
\end{equation}
%-----------------
The factor $\theta(E)$ emphasizes that the line shape 
was assumed to be zero below the $D^{*0} \bar D^0$ threshold.
This reflects the incorrect assumption that $D^0 \bar D^0 \pi^0$ 
events can come only from the production of 
$D^{*0} \bar D^0$ or $D^{0} \bar D^{*0}$ above the threshold 
followed by the decay of $D^{*0}$ or $\bar D^{*0}$
and not from the decay of a bound state below the threshold.
However the mass of the $X(3872)$ is about 7~MeV above the 
$D^0 \bar D^0 \pi^0$ threshold, so there is plenty of phase space 
for the decay of this bound state into $D^0 \bar D^0 \pi^0$.  
Moreover the $X(3872)$ spends most of its time in a configuration 
in which the charm mesons have large separation, so the 
$D^{*0}$ or $\bar D^{*0}$ in the bound state can decay 
almost as if they were free particles.
The conceptual error in Ref.~\cite{Hanhart:2007yq} 
was pointed out in Ref.~\cite{Braaten:2007dw},
and an analysis that takes into account the decay of the bound state
was carried out. 

Zhang, Meng and Zheng have recently carried out an updated analysis 
\cite{Zhang:2009bv} of the recent data from Babar and Belle using 
essentially the same line shapes as 
Hanhart et al.  They repeated the conceptual error  
of Ref.~\cite{Hanhart:2007yq} by taking the line shape in 
$D^{*0} \bar D^0$ to be given by Eq.~(\ref{fHKKN-DDpi}),
which does not take into account $D^0 \bar D^0 \pi^0$ events
produced by decays of the bound state.
They determined the location of the poles
in the energy $E$ for the scattering amplitude $f_{\rm HKKN}(E)$ 
in Eq.~(\ref{f-Flatte}).  All of their fits had one pole 
for which the real and imaginary parts of $E$ were less than 1~MeV.  
This is the pole associated 
with the S-wave threshold resonance.  Their fits also had a second pole 
on a different Riemann sheet of the complex energy $E$ 
whose absolute value was significantly larger than 1~MeV.
This pole is an artifact of the scattering amplitude
in Eq.~(\ref{f-Flatte}) and has no physical significance.

\section{Summary}
\label{sec:summary}
We have carried out an analysis of the line shapes of the $X(3872)$ 
in the $J/\psi \, \pi^+ \pi^-$ and $D^0 \bar D^0 \pi^0$ decay channels
using the most recent data from the Babar and Belle Collaborations.
For the signal, we used the line shapes of an S-wave threshold 
resonance, which differ in several crucial respects from the 
conventional Breit-Wigner resonance.   We took 
into account the experimental resolution in the energy distributions
using Gaussian smearing functions.  In the case of the
$D^0 \bar D^0 \pi^0$ channel, we also took into account the assumption 
in the experimental analyses that $D^0 \bar D^0 \pi^0$ events
near the $D^{*0} \bar D^0$ threshold
come from $D^{*0} \bar D^0$ and $D^{0} \bar D^{*0}$.

The parameters for the S-wave threshold resonance are 
the real and imaginary parts of the inverse scattering length
$\gamma_{\rm re}+ i \gamma_{\rm im}$ and a normalization factor 
$({\cal B} {\cal B})_F$ that depends on the decay channel $F$.
A characteristic feature of an S-wave threshold resonance
is that its line shapes are not integrable functions of the energy.
One consequence is that the product $({\cal B} {\cal B})_F$
of the branching fractions for the production of the
resonance and its decay into the final state $F$ depends on the
prescription used to define it.  Our prescription for 
$({\cal B} {\cal B})_{J/\psi \, \pi^+ \pi^-}$
is specified by the analytic expression for the line shape
in Eq.~(\ref{Eq:dGdE}).  Our prescription for 
$({\cal B} {\cal B})_{D^0 \bar D^0 \pi^0}$
is specified by the analytic expression for the line shape
in Eq.~(\ref{Eq:dGamDDpi}).
The parameters for our fits to the Babar and Belle data
in the $J/\psi \, \pi^+ \pi^-$ and $D^0 \bar D^0 \pi^0$ decay channels
are given in Tables~\ref{tab:psipipi} and \ref{tab:DDpi}.

Because the line shape of an S-wave threshold resonance is not an 
integrable function of the energy, a prescription is required 
to define the binding energy and the width of the $X(3872)$.
Our prescriptions for the binding energy $E_X$ and the width $\Gamma_X$ 
are that the pole in the amplitude as a function 
of the complex energy $E$ are at $-E_X - i \Gamma_X/2$.
Given the values of $\gamma_{\rm re}$ and $\gamma_{\rm im}$,
$E_X$ and $\Gamma_X$ can be calculated using 
Eqs.~(\ref{Eq:EX-GamX}).  
An alternative pair of variables that can in principle be measured 
directly are the position $E_{\rm max}$ of the peak in the line shape
and its full width at half-maximum $\Gamma_{\rm fwhm}$.
Given the values of $\gamma_{\rm re}$ and $\gamma_{\rm im}$,
$E_{\rm max}$ and $\Gamma_{\rm fwhm}$ can be calculated 
by solving Eqs.~(\ref{Eq:eqEmax}) and (\ref{Eq:eqGammafwhm}).
The values of $-E_X$ , $\Gamma_X$, $E_{\rm max}$, and $\Gamma_{\rm fwhm}$ 
for our fits to the Babar and Belle data
in the $J/\psi \, \pi^+ \pi^-$ and $D^0 \bar D^0 \pi^0$ decay channels
are listed in Tables~\ref{tab:psipipi} and \ref{tab:DDpi}.

We carried out two fits to each of the data sets from the
Belle and Babar Collaborations, one with $\gamma_{\rm im} = 0$
and one with $\gamma_{\rm im}$ as a fitting parameter.
The best fits to the smeared line shapes in the $J/\psi \, \pi^+ \pi^-$  
decay channel are shown in Figs.~\ref{Fig:psipipiBabar} 
and \ref{Fig:psipipiBelle}.
The best fits to the smeared $D^{*0} \bar D^0$ energy distributions 
are shown in Figs.~\ref{Fig:DDpiBabar} and \ref{Fig:DDpiBelle}.
The line shapes in the $J/\psi \, \pi^+ \pi^-$ 
and $D^0 \bar D^0 \pi^0$ decay channels and the $D^{*0} \bar D^0$ 
energy distributions for three of the best fits
are shown in Figs.~\ref{Fig:LSreal}, \ref{Fig:LScomplex}, and
\ref{Fig:LSfit}.  The $D^0 \bar D^0 \pi^0$ line shape has a peak below the
$D^{*0} \bar D^0$ threshold whose position and width are
close to those for the $J/\psi \, \pi^+ \pi^-$ line shape. 
It is this peak that should be identified as the $X(3872)$ resonance.
The $D^{*0} \bar D^0$ energy distribution, which vanishes below the 
threshold, has a peak above the threshold whose width 
is considerably larger than the width of the $X(3872)$ resonance.
Thus measurements of the position and width of the peak 
in the $D^{*0} \bar D^0$ invariant mass distribution should not be 
interpreted as measurements of the
mass and width of the $X(3872)$.

In our analyses of the $D^{*0} \bar D^0$ energy distributions
measured by the Babar and Belle Collaborations,
we took into account the assumption that $D^0 \bar D^0 \pi^0$ events
near the $D^{*0} \bar D^0$ threshold
come from $D^{*0} \bar D^0$ and $D^{0} \bar D^{*0}$.
Even though the $D^{*0} \bar D^0$ energy distributions
vanish below the $D^{*0} \bar D^0$ threshold, our analyses 
of these distributions gave values for
the position of the $X(3872)$ resonance that were below the threshold.
In our analyses with $\gamma_{\rm im}$ as a fitting parameter,
the best fits were for $\gamma_{\rm im} = 0$, which is the minimum 
physical value.  In contrast, the best fits to the Babar and 
Belle data on $J/\psi \, \pi^+ \pi^-$ gave 
$\gamma_{\rm im} = 15.5$~MeV and 12.0~MeV, respectively.
The preference for the value $\gamma_{\rm im} = 0$ in the fit
to the  $D^0 \bar D^0 \pi^0$ data could be an artifact of 
the simple model for the $D^{*0} \bar D^0$ energy resolution
given in Eqs.~(\ref{Eq:RE'}) and (\ref{smear:DDpi}).
Because this model is questionable, we regard our analyses of the 
$D^0 \bar D^0 \pi^0$ data as only illustrative.
They are no substitute for analyses by the experimental
collaborations that take all the correlated errors properly 
into account.  In a careful analysis, it would be better to use the
differential decay distribution in Eq.~(\ref{Eq:dGamDDpi})
instead of our analytic expression for the line shape in 
Eq.~(\ref{Eq:dGdEDDpi}).  Finally an analysis of the
$D^0 \bar D^0 \pi^0$ decay channel similar to the
original Belle analysis in Ref.~\cite{Gokhroo:2006bt}
would be preferable to one in which $D^0 \bar D^0 \pi^0$ events
near the $D^{*0} \bar D^0$ threshold
are interpreted as $D^{*0} \bar D^0$ or $D^{0} \bar D^{*0}$.
If such an analysis gave resonance parameters for the 
$X(3872)$ that are close to those from analyses of the 
$J/\psi \, \pi^+ \pi^-$ channel, it would go a long way 
towards solidifying a consensus in the high energy physics community 
on the nature of the $X(3872)$.

\begin{acknowledgments}
This research was supported in part by the Department of Energy
under grant DE-FG02-91-ER40690.
\end{acknowledgments}



\begin{thebibliography}{}

% X(3872) discovery
%\cite{Choi:2003ue}
\bibitem{Choi:2003ue}
  S.K.~Choi {\it et al.}  [Belle Collaboration],
%  ``Observation of a new narrow charmonium state in exclusive 
%    $B^\pm \to K^\pm  \pi^+ \pi^- J/\psi$ decays,''
  Phys.\ Rev.\ Lett.\  {\bf 91}, 262001 (2003)
  [arXiv:hep-ex/0309032].
  %%CITATION = PRLTA,91,262001;%%

%\cite{Aubert:2008gu}
\bibitem{Aubert:2008gu}
  B.~Aubert {\it et al.}  [BABAR Collaboration],
%  ``A study of $B \to X(3872) K$, with $X(3872) \to J/\psi \pi^+ \pi^-$,''
  Phys.\ Rev.\  D {\bf 77}, 111101 (2008)
  [arXiv:0803.2838 [hep-ex]].

%\cite{Belle:2008te}
\bibitem{Belle:2008te}
 I.~Adachi {\it et al.}  [Belle Collaboration],
%  ``Study of $X(3872)$ in $B$ meson decays,''
  arXiv:0809.1224 [hep-ex].
  
%\cite{Aaltonen:2009vj}
\bibitem{Aaltonen:2009vj}
  T.~Aaltonen {\it et al.}  [CDF Collaboration],
%  ``Precision Measurement of the $X(3872)$ Mass in $J/\psi \pi^+ \pi^-$ Decays,''
  arXiv:0906.5218 [hep-ex].

%\cite{Abe:2005ix}
\bibitem{Abe:2005ix}
  K.~Abe {\it et al.},
%  ``Evidence for $X(3872) \to \gamma J/\psi$ and the sub-threshold decay 
%  $X(3872) \to \omega J/\psi$,''
  arXiv:hep-ex/0505037.

%\cite{Aubert:2006aj}
\bibitem{Aubert:2006aj}
  B.~Aubert {\it et al.}  [BABAR Collaboration],
%  ``Search for $B^{+} \to X(3872) K^{+}$, $X_{3872} \to J/\psi \gamma$,''
  Phys.\ Rev.\  D {\bf 74}, 071101 (2006)
  [arXiv:hep-ex/0607050].

%\cite{Babar:2008rn}
\bibitem{Babar:2008rn}
  B.~Fulsom {\it et al.}  [BABAR Collaboration],
%  ``Evidence for $X(3872) \to \psi(2S) \gamma$ in $B^\pm \to X(3872) K^\pm$ 
%    decays, and a study of $B \to c \bar c \gamma K$,''
  arXiv:0809.0042 [hep-ex].

%\cite{Abe:2005iya}
\bibitem{Abe:2005iya}
  K.~Abe {\it et al.},
%  ``Experimental constraints on the possible $J(PC)$ quantum numbers of the
%  $X(3872)$,''
  arXiv:hep-ex/0505038.

%\cite{Abulencia:2006ma}
\bibitem{Abulencia:2006ma}
  A.~Abulencia {\it et al.}  [CDF Collaboration],
%  ``Analysis of the quantum numbers $J(PC)$ of the $X(3872)$,''
  Phys.\ Rev.\ Lett.\  {\bf 98}, 132002 (2007)
  [arXiv:hep-ex/0612053].
  
%\cite{Gokhroo:2006bt}
\bibitem{Gokhroo:2006bt}
  G.~Gokhroo {\it et al.},
%  ``Observation of a near-threshold $D^0 \bar D^0 \pi^0$ enhancement in 
%    $B \to D^0 \bar D^0 \pi^0 K$ decay,''
  Phys.\ Rev.\ Lett.\  {\bf 97}, 162002 (2006).
  [arXiv:hep-ex/0606055].

%\cite{Babar:2007rva}
\bibitem{Babar:2007rva}
  B.~Aubert {\it et al.}  [BABAR Collaboration],
%  ``Study of Resonances in Exclusive $B$ Decays to $\bar D^{(*)} D^{(*)} K$,''
  arXiv:0708.1565 [hep-ex].

%\cite{Belle:2008su}
\bibitem{Belle:2008su}
I.~Adachi {\it et al.}  [Belle Collaboration],
%  ``Study of the $B \to X(3872)(D^{*0} \bar{D}^0) K$ decay,''
  arXiv:0810.0358 [hep-ex].

%\cite{Braaten:2004rn}
\bibitem{Braaten:2004rn}
  E.~Braaten and H.W.~Hammer,
%  ``Universality in Few-body Systems with Large Scattering Length,''
  Phys.\ Rept.\  {\bf 428}, 259 (2006).
  [arXiv:cond-mat/0410417].

%\cite{Amsler:2008zzb}
\bibitem{Amsler:2008zzb}
  C.~Amsler {\it et al.}  [Particle Data Group],
%  ``Review of particle physics,''
  Phys.\ Lett.\  B {\bf 667}, 1 (2008).
 
%\cite{Hanhart:2007yq}
\bibitem{Hanhart:2007yq}
  C.~Hanhart, Yu.S.~Kalashnikova, A.E.~Kudryavtsev, and A.V.~Nefediev,
%  ``Reconciling the $X(3872)$ with the near-threshold enhancement in
%    the $D^0 \bar{D}^{*0}$ final state,''
  Phys.\ Rev.\  D {\bf 76}, 034007 (2007)
  [arXiv:0704.0605 [hep-ph]].

%\cite{Voloshin:2007hh}
\bibitem{Voloshin:2007hh}
  M.B.~Voloshin,
%  ``Isospin properties of the $X$ state near the $D {\bar D}^{*}$ threshold,''
  Phys.\ Rev.\  D {\bf 76}, 014007 (2007)
  [arXiv:0704.3029 [hep-ph]].

%\cite{Zhang:2009bv}
\bibitem{Zhang:2009bv}
  O.~Zhang, C.~Meng and H.~Q.~Zheng,
%  ``Ambiversion of $X(3872)$,''
  arXiv:0901.1553 [hep-ph].
  
%\cite{Braaten:2007dw}
\bibitem{Braaten:2007dw}
  E.~Braaten and M.~Lu,
%  ``Line Shapes of the $X(3872)$,''
  Phys.\ Rev.\  D {\bf 76}, 094028 (2007)
  [arXiv:0709.2697 [hep-ph]].

%\cite{Voloshin:2003nt}
\bibitem{Voloshin:2003nt}
  M.~B.~Voloshin,
%  ``Interference and binding effects in decays of possible molecular component
%  of X(3872),''
  Phys.\ Lett.\  B {\bf 579}, 316 (2004)
  [arXiv:hep-ph/0309307].

%\cite{Braaten:2007ft}
\bibitem{Braaten:2007ft}
  E.~Braaten and M.~Lu,
%  ``The Effects of Charged Charm Mesons on the Line Shapes of the X(3872),''
  Phys.\ Rev.\  D {\bf 77}, 014029 (2008)
  [arXiv:0710.5482 [hep-ph]].

%\cite{Uehara:2005qd}
\bibitem{Uehara:2005qd}
  S.~Uehara {\it et al.}  [Belle Collaboration],
%  ``Observation of a $\chi'_{c2}$ candidate in $\gamma \gamma \to D \bar D$  
%  production at Belle,''
  Phys.\ Rev.\ Lett.\  {\bf 96}, 082003 (2006)
  [arXiv:hep-ex/0512035].

%\cite{Barnes:2005pb}
\bibitem{Barnes:2005pb}
  T.~Barnes, S.~Godfrey and E.~S.~Swanson,
%  ``Higher Charmonia,''
  Phys.\ Rev.\  D {\bf 72}, 054026 (2005)
  [arXiv:hep-ph/0505002].

%\cite{Eichten:2005ga}
\bibitem{Eichten:2005ga}
  E.~J.~Eichten, K.~Lane and C.~Quigg,
%  ``New states above charm threshold,''
  Phys.\ Rev.\  D {\bf 73}, 014014 (2006)
  [Erratum-ibid.\  D {\bf 73}, 079903 (2006)]
  [arXiv:hep-ph/0511179].

%\cite{Flatte:1976xu}
\bibitem{Flatte:1976xu}
  S.~M.~Flatte,
%  ``Coupled-channel Analysis of the $\pi \eta$ and $K \bar K$ Systems 
%    near $K \bar K$ Threshold,''
  Phys.\ Lett.\  B {\bf 63}, 224 (1976).
  
\end{thebibliography}
\end{document}